\documentclass{emulateapj}

\newcommand{\Lsol}{\mbox{$L_\odot$}}
\newcommand{\Msol}{\mbox{$M_\odot$}}

\newcommand{\asec}{\mbox{$''$}}
\newcommand{\kms}{\mbox{km s$^{-1}$}}
\newcommand{\perkms}{\mbox{(km s$^{-1}$)$^{-1}$}}
\newcommand{\kmsperpc}{\mbox{km s$^{-1}$ pc$^{-1}$}}

\newcommand{\perbeam}{\mbox{beam$^{-1}$}}
\newcommand{\persquarecm}{\mbox{cm$^{-2}$}}
\newcommand{\percubiccm}{\mbox{cm$^{-3}$}}

\newcommand{\persquarearcsec}{\mbox{arcsec$^{-2}$}}
\newcommand{\perpc}{\mbox{pc$^{-1}$}}
\newcommand{\persquarepc}{\mbox{pc$^{-2}$}}
\newcommand{\percubicpc}{\mbox{pc$^{-3}$}}

\newcommand{\plus}{\mbox{$+$}}
\newcommand{\minus}{\mbox{$-$}}

\newcommand{\uv}{\mbox{$u$--$v$}}

\newcommand{\about}{\mbox{$\sim$}}
\newcommand{\hr}{\mbox{$^{\rm h}$}}
\newcommand{\mn}{\mbox{$^{\rm m}$}}

\newcommand{\HH}{\mbox{H$_2$}}

\newcommand{\ammonia}{\mbox{NH$_3$}}

\newcommand{\HCOplus}{\mbox{HCO$^{+}$}}
\newcommand{\nd}{\nodata}
\newcommand{\tnm}[1]{\tablenotemark{#1}}
\newcommand{\Lir}{\mbox{$L_{\rm 8-1000\; \mu m}$}}
\newcommand{\Tb}{\mbox{$T_{\rm b}$}}

\newcommand{\KRJ}{\mbox{$\rm K_{RJ}$}}

\newcommand{\unitofX}{\mbox{cm$^{-2}$ (\KRJ\ km s$^{-1}$)$^{-1}$}}
\newcommand{\unitofkappa}{\mbox{(g cm$^{-2}$)$^{-1}$}}
\newcommand{\taum}{\mbox{$\tau_{\rm m}$}}
\newcommand{\taud}{\mbox{$\tau_{\rm d}$}}
\newcommand{\Tcmb}{\mbox{$T_{\rm CMB}$}}
\newcommand{\Tline}{\mbox{$T_{\rm b, excess}^{\rm line}$}}
\newcommand{\Tcont}{\mbox{$T_{\rm b, excess}^{\rm cont.}$}}
\newcommand{\Weq}{\mbox{$W_{\rm eq}$}}

\newcommand{\Nhh}{\mbox{$N_{\rm H_{2}}$}}
\newcommand{\DeltaV}{\mbox{$\Delta V$}}
\newcommand{\altdef}{\mbox{$\equiv$}}
\newcommand{\citest}[1]{\citeauthor*{#1}}
\newcommand{\citesp}[1]{(\citeauthor*{#1})}

\shorttitle{860 \micron\ Imaging of Arp 220}
\shortauthors{SAKAMOTO et al.}
\slugcomment{Accepted for publication in ApJ. 2008-05-30}

\begin{document}
\title{SMA Imaging of CO(3--2) Line and 860 \micron\ Continuum of Arp 220 : \\
Tracing the Spatial Distribution of Luminosity 
}

\author{Kazushi Sakamoto\altaffilmark{1,2,7}, 
Junzhi Wang\altaffilmark{3,4},
Martina C. Wiedner\altaffilmark{3,5},
Zhong Wang\altaffilmark{3}, \\
Alison B. Peck\altaffilmark{1,6},
Qizhou Zhang\altaffilmark{3}, 
Glen R. Petitpas\altaffilmark{1}, \\  
Paul T. P. Ho\altaffilmark{3,7},
and
David J. Wilner\altaffilmark{3}
}
\altaffiltext{1}{Harvard-Smithsonian Center for Astrophysics,
Submillimeter Array, 645, N. A'ohoku Place, Hilo, HI 96720}
\altaffiltext{2}{National Astronomical Observatory of Japan,
Mitaka, Tokyo 181-8588, Japan  } 
\altaffiltext{3}{Harvard-Smithsonian Center for Astrophysics,
60 Garden Street, Cambridge, MA 02138}
\altaffiltext{4}{Department of Astronomy, Nanjing University,   Nanjing, 210093, China}
\altaffiltext{5}{I. Physikalisches Institut, Universit\"at zu K\"oln, Z\"ulpicher Str. 77,
50937 K\"oln, Germany}
\altaffiltext{6}{ALMA JAO, Av. El Golf 40 - Piso 18, Las Condes, Santiago, Chile}
\altaffiltext{7}{Academia Sinica, Institute of Astronomy and Astrophysics, 
P.O. Box 23-141, Taipei 106, Taiwan}

\begin{abstract} 
We used the Submillimeter Array (SMA) to image 860 \micron\ continuum
and CO(3--2) line emission in the ultraluminous merging galaxy Arp 220, achieving
a resolution of 0\farcs23 (80 pc) for the continuum and 0\farcs33 (120 pc) for the line.
The CO emission peaks around the two merger nuclei 
with a velocity signature of gas rotation around each nucleus,
and is also detected in a kpc-size disk encompassing the binary nucleus.
The dust continuum, in contrast, is mostly from the two nuclei. 
The beam-averaged brightness temperature of both line and continuum emission 
exceeds 50 K at and around the nuclei,
revealing the presence of warm molecular gas and dust.
The dust emission morphologically agrees with the distribution of radio supernova features 
in the east nucleus, as expected when a starburst heats the nucleus.
In the brighter west nucleus, however, the submillimeter dust emission is more compact than 
the supernova distribution.
The 860 \micron\ core, after deconvolution, has a size of 50--80 pc, 
consistent with recent 1.3 mm observations,
and  a peak brightness temperature of (0.9--1.6)$\times 10^2$ K.
Its bolometric luminosity is at least $2\times10^{11}$ \Lsol\ and could be  \about$10^{12}$ \Lsol\
depending on source structure and 860 \micron\ opacity, which we estimate to be
of the order of  $\tau_{860} \sim 1$ (i.e.,  $N_{\rm H_{2}} \sim 10^{25}$ \persquarecm).
The starbursting west nuclear disk must have in its center 
a dust enshrouded AGN 
or
a very young starburst equivalent to hundreds of super star clusters.
Further spatial mapping of bolometric luminosity through submillimeter imaging is a promising
way to identify the heavily obscured heating sources in Arp 220 and other luminous infrared galaxies.
\end{abstract}

\keywords{ 
        galaxies: ISM ---
        galaxies: starburst ---
        quasars: general ---
        galaxies: evolution---        
        galaxies: individual (Arp 220) 
       }

\section{Introduction}  
\label{s.introduction}
\object{Arp 220} is the nearest ultraluminous infrared galaxy (ULIRG) and is an advanced merger
($\Lir=10^{12.2} \Lsol$; $D=75$ Mpc; 1\arcsec = 361 pc; see Table \ref{t.galparm}).
Its vast luminosity is radiated almost entirely  between 
infrared and millimeter wavelengths as thermal dust emission. 
This is because the source, an extreme starburst, a quasar-class AGN(s), or both,
is deeply embedded in dust  within the central few kpc \citep{Soifer84,Joy86}.
This merger has been a prime target to study the mechanism to generate the 
high luminosity of ULIRGs and to study galaxy evolution through major mergers 
\citep{Sanders96, Genzel98}. 
Detailed studies of Arp 220 as well as other local ULIRGs may also help us to understand
submillimeter-detected luminous galaxies at high redshift \citep[e.g.,][]{Coppin06}, because the 
latter may be scaled-up versions of the former \citep{Tacconi06}.

The Arp 220 system has two nuclei, presumably from the merger progenitors, with a projected 
separation of about 350 pc 
\citep{Norris88,Graham90,Clements02}.
Millimeter CO observations reveal a large amount of molecular gas 
(several $10^{9}$ \Msol) in the central few kpc with peaks at the nuclei
(\citealt{Scoville91};
 \citealt{Scoville97} (\altdef \citest{Scoville97});
 \citealt{Downes98};
 \citealt{Sakamoto99} (\altdef \citest{Sakamoto99});
 \citealt{Downes07} (\altdef \citest{Downes07})).
Sub-arcsecond resolution CO and \ion{H}{1} observations show that 
a small gas disk ($r\sim 50$ pc) rotates around each nucleus 
and that the two nuclear disks, counterrotating with respect to each other, 
are encompassed by a larger outer disk 
(\citest{Sakamoto99};
 \citealt{Mundell01} (\altdef \citest{Mundell01}), in which one of the
nuclear disks was modeled as a part of the outer disk.)
The two nuclei dominate the dust emission at 1.3 mm \citesp{Sakamoto99} and
in mid-IR up to 25 \micron\ \citep[$\equiv$\citest{Soifer99}]{Soifer99}, 
suggesting them to be a major source of the luminosity.
Indeed, vigorous starburst activity in the nuclear disks was found
from a concentration of radio supernovae and young supernova remnants
within about 50 pc (\about0\farcs2) of each nucleus 
and from radio recombination lines in the nuclear disks \citep{Smith98,Rovilos05,Lonsdale06,Anan00,Rodriguez05}.
The presence of an AGN in Arp 220 has not been proven nor rejected based on X-ray observations
because of high obscuration \citep{Iwasawa01,Iwasawa05,Clements02}.

Arp 220 appears to represent a short luminous stage in merger evolution right before 
the coalescence of two nuclei, because there are luminous merging galaxies having similar structure.
The advanced merger \object{NGC 3256} ($\Lir=10^{11.6} \Lsol$)
has twin gas peaks at two nuclei, possible rotation of the gas around each nucleus, and a large ($>$ kpc) gas disk
surrounding the central region \citep{Sakamoto06.n3256}. 
\object{NGC 6240} ($\Lir=10^{11.9} \Lsol$) also has two nuclei within 1 kpc on the sky, with each having a rotating
stellar core and being an AGN \citep{Tecza00, Komossa03}.
Studying Arp 220 is particularly important to understand this class of objects.

Accurate spatial distribution of the luminosity should tell us much about the hidden energy source(s) in Arp 220.
The dichotomy of the nuclei (with the nuclear disks) and the outer disk has been proven useful
for this approach.
\citest{Soifer99} showed using mid-IR imaging 
and the spectral energy distribution (SED) from infrared to millimeter wavelengths that
the two nuclei  of \about0\farcs3 (\about100 pc) extent have at least 40\% of the total luminosity of Arp 220.
They also pointed out that the outer disk, or a kpc-size extended component around the nuclei, 
could be another major component of the luminosity,
contributing to the SED mainly at around 60--100 \micron\
but too faint in continuum to be detected at mid-IR and millimeter wavelengths.
Further mid-IR spectroscopy and SED modeling suggested, however,
that the less-obscured outer disk has a moderate starburst with up to 15\%
of the total luminosity,
leaving a large majority of the luminosity to the heavily obscured nuclei
\citep{Soifer02,Spoon04,Gonzalez04}. 
The heavily embedded and compact source of luminosity was also suggested by \citet{Dudley97} from
mid-IR silicate absorption.
Most of the luminosity in the nuclei is attributed to starburst by the groups that observed 
the radio supernovae and recombination lines.
Still, there have been arguments for an AGN(s) 
\citep[both based on (sub)millimeter line ratios favoring X-ray dominated regions]{Aalto07,Imanishi07} 
and even for its dominance in luminosity
(\citealt{Haas01}, based on deficit of 7.7 \micron\ PAH with respect to 850 \micron\ continuum; 
see \citealt{Spoon04} for caveats).

Recently, \citest{Downes07} obtained from their 0\farcs3-resolution observations
of 1.3 mm continuum 
a deconvolved size of $0\farcs19 \times 0\farcs13$ and
a deconvolved peak brightness-temperature of 90 K for the western nucleus.
They attributed to the compact source a probable
total luminosity of the nucleus, $9\times10^{11}$ \Lsol\ in the nucleus-dominated model of \citest{Soifer99},
and estimated the intrinsic dust temperature to be 170 K 
and the 1.3 mm opacity to be \about0.7. 
They argued that the heating source must be an AGN on the basis of the high luminosity surface density.
Although this model is possible, the minimum luminosity allowed by the 1.3 mm data is
$3 \times 10^{10}$ \Lsol\  in the blackbody limit; 
it is $1 \times 10^{11}$ \Lsol\ if the 1.3 mm opacity is 1.3
as can be estimated from a power-law index of $\beta=2$ for dust emissivity and 
the 2.6 mm dust flux in \citest{Downes07}.
Thus the majority of the luminosity in the west nucleus may come from the starburst in its
nuclear disk rather than from its continuum core, 
and the source for the central luminosity may also be massive stars.
Our quest for the accurate spatial distribution and the source of the luminosity in the merger is
therefore not yet over.

In this paper, we report 0\farcs2--0\farcs3 resolution imaging of Arp 220 carried out with the
SMA in the submillimeter CO(3--2) line and 860 \micron\ continuum.
In submillimeter, where dust opacity is higher than in millimeter, we can better trace
warm dust and better estimate the dust temperature and luminosity with less uncertainty from
opacity correction.
This is the first spatially-resolved imaging of the merger at a submillimeter wavelength
and the second submillimeter interferometric observations of the galaxy 
following the original single-baseline experiment with the JCMT and CSO  \citep{Wiedner02}.
The SMA provided us with comparable or higher resolutions than previous
millimeter observations.
Our observations are described in \S \ref{s.obs} and data reduction in \S \ref{s.reduction}.
The results are presented in \S \ref{s.result.co} for the CO line and in \S \ref{s.result.continuum} for continuum. 
We analyze the data for the dust opacity and temperature,
and set a larger lower limit to the luminosity in the core of the west nucleus
than the aforementioned limit from millimeter data by about an order of magnitude.
The continuum data are also compared with the supernova distribution, 
and the line data are analyzed for gas dynamics and ISM properties in and around the nuclei.
We discuss the properties of the hot nuclei and their energy source(s) in \S \ref{s.discuss}, 
showing that some luminosity-related parameters of the west nucleus are close to the highest
for a starburst and that submillimeter observations may soon resolve the energy source issue.
Our finding are summarized in \S \ref{s.summary}.

\section{SMA Observations} 
\label{s.obs}

The Submillimeter Array 
(SMA)\footnote{
The Submillimeter Array is a joint
project between the Smithsonian Astrophysical Observatory and the
Academia Sinica Institute of Astronomy and Astrophysics, and is
funded by the Smithsonian Institution and the Academia Sinica.
} on Mauna Kea, Hawaii, was used to observe Arp 220 in 2004--2006. 
The array has eight 6-meter antennas, cryogenically cooled SIS receivers,
and a digital correlator \citep{Ho04}.
The parameters and the log of the observations are in Tables \ref{t.obsparam} and \ref{t.obslog}, respectively.

The redshifted CO(3--2) line was placed in the center of the lower side band (LSB) and the continuum
data were obtained from the LSB line-free channels and from the upper side band (USB).
The antennas were pointed to the center of the merger.  
Their primary beams are approximately Gaussian with a full width half maximum (FWHM)
of 35\arcsec\ in the LSB.
The correlator was configured for 3.25 MHz resolution and 2 GHz band width in each sideband.
These correspond to 2.8 \kms\ and 1734 \kms, respectively, for the CO(3--2) line.
Data in both sidebands were recorded through sideband separation using 90\arcdeg\ phase switching.

In each night we observed
gain calibrator(s), passband calibrator(s), the target galaxy, and, in
many cases, the flux calibrator Uranus.
Either J1613+342 or J1635+381 or both were observed every 20--30 min during each track
to calibrate complex gain.
Their positions used for our observations agree to 3 mas with those from VLBI measurements in
\citet{Ma98}, tying our astrometry to the International Celestial Reference System.
Passband calibration was made by observing planets and bright quasars such as
3C279 and 3C454.3.
Pointing was checked before and during each track on bright quasars.

Observations were made on 10 nights but 6 nights were lost to poor seeing or various
instrumental problems and 1 night was useable only for flux calibration (track No. 5 in Table \ref{t.obslog}). 
The atmosphere was relatively transparent during the four (partially) useable nights;
the zenith opacity measured at 225 GHz was below 0.1 
and hence the 345 GHz opacity  below 0.4 \citep[see][for the conversion]{Matsuo98}.

The observations in multiple antenna configurations 
gave us projected baselines for the galaxy in the range of 10.2 m -- 508.1 m after flagging bad data.
The range corresponds to the fringe spacing of 18\arcsec\ -- 0\farcs35 at 345 GHz.
Our longest projected baseline almost matches the longest physical baseline of the SMA (508.9 m)
because Arp 220 transits only 4\arcdeg\ from the zenith on Mauna Kea.

\section{Data reduction} 
\label{s.reduction}
We reduced our data in three steps.
First, the standard gain and passband calibrations were made with MIR
\citep[see][for its original form]{Scoville93a}.
Atmospheric attenuation was also corrected at this point using the system temperature 
monitored throughout the observations.
Next, images were made from the calibrated visibilities using MIRIAD \citep{Sault95}.
Velocity channels were binned to 30 \kms\ in this step.
We made two sets of images, at lower and higher resolution, 
by using different weighting and selection of the visibilities.
The former has a resolution of 0\farcs5, while the latter
has a resolution of 0\farcs33 for the CO line and 0\farcs23 for the continuum.
The high-resolution continuum image used only track No. 6 taken in the most extended array configuration,
while the rest of the images used tracks No. 3, 6, and 7.
Finally, the images and data cubes were analyzed using the NRAO AIPS \citep{Bridle94}.
Details of the data reduction are in \S \ref{s.red.flux} -- \S \ref{s.red.comparison}.

Throughout this paper, velocities are defined in the radio convention,
 $v \equiv c(1- f_{\rm obs}/f_{\rm rest})$, with respect to the local standard of rest (LSR).
All the coordinates are in the J2000.0 system.
The brightness temperature in this paper is the excess temperature above the cosmic background,
which is only 3 mJy \persquarearcsec\ at 345 GHz.
Brightness temperatures are calculated with the Planck function unless
the Rayleigh-Jeans approximation is noted or, equivalently, the unit is \KRJ.

\subsection{Flux Calibration} 
\label{s.red.flux}
The flux scale in our data is based on observations of Uranus and its brightness temperature 
given by \citet{Griffin93}. 
Flux densities of our gain calibrators were determined by comparison with the planet.
We separately derived the quasar flux density in the two sidebands and averaged the
results to obtain the flux density at 340--350 GHz, effectively
ignoring the small spectral slope of the quasars 
\citep[typically $\alpha \sim -0.5$ for $S_{\nu} \propto \nu^{\alpha}$;][]{Gear94}.

The uncertainty of the flux scale of our data was estimated to be about $\pm$15\%
by examining relative flux scale between our tracks. 
For this, the flux of Arp 220 in each sideband was measured in each track 
from the common baselines in the common ranges of  hour angle 
in order to ensure similar \uv\ sampling. 
The flux data agreed within 15\%  between the tracks having planet-based flux calibration.
This method was also used to establish the flux scale in track No. 3, which had poor seeing
during Uranus observations.

We safely ignored the response pattern of the primary beam in all flux and intensity measurements
in our data because the emission of the merger is compact.
Emission 2\arcsec\ from the center is attenuated by only 1\%.

\subsection{Atmospheric Seeing and Astrometry} 
\label{s.red.seeing}
We made a test to determine the point spread function and to estimate our astrometric accuracy.
The data from the very-extended configuration (track No. 6) were calibrated in the test 
using only one of the two gain calibrators.
When calibrating the quasar J1635+381 with  J1613+342, it had 
a finite deconvolved size of 0\farcs09 in FWHM.
We attribute this to atmospheric seeing and  imperfect calibration. 
(To be precise, the 0\farcs09 is not exactly the size of the so-called `seeing disk'  
because the atmospheric decorrelation common to the two quasars is calibrated out.)
Astrometric accuracy turned out to be  0\farcs03 between the two quasars.
Our astrometric error for Arp 220 may be larger than this because the galaxy
is 14\arcdeg\ and 19\arcdeg\ from the quasars while the quasars are only 6\arcdeg\ apart.
We used both quasars in the gain calibration for the data presented below.

\subsection{Continuum} 
\label{s.red.continuum}

\subsubsection{Choice of Spectrometer Channels for Continuum}
The continuum data of Arp 220 were obtained by integrating line-free channels,
but the determination of the line-free channels needed some care.
This is because the line emission may have a broad and low-intensity wing, and also because
there may be weak lines adjacent to the target CO line.
Recent detections of \about 1800 \kms-wide absorption of \ammonia\ toward
Arp 220 \citep{Takano05} and of high-velocity CO emission in the similar luminous merger 
NGC 3256 \citep{Sakamoto06.n3256} gave us reason to be cautious about 
possible high-velocity gas.

We adopted as continuum channels those outside of the velocity range of 4800--5900 \kms\ 
after examining the CO(2--1) spectrum of Arp 220 shown in Fig. \ref{f.co2-1spectrum}.
The spectrum is from the dual band (690 GHz and 230 GHz) SMA observations  on 2005 March 2nd
\citep{Matsushita07}; the array was in the compact configuration.
The CO(2--1) spectrum has a higher signal-to-noise ratio and a wider velocity coverage
than our CO(3--2) data, and it does not show high-velocity emission outside the
abovementioned velocity range.
We inferred from this that 
CO(3--2) emission from any high-velocity gas should be below our sensitivity limit
unless the CO excitation of the gas (if it exists) were drastically different from 
that in the main gas component of the merger.

The continuum data used for the lower resolution (0\farcs5) image were
from both the lower and upper sidebands and were from the same
line-free spectrometer channels.
This choice of the USB channels was a precaution against any leakage of the LSB CO line to the USB 
due to imperfect sideband separation.
An error in the SMA software had caused a leakage of about $-13$ dB 
between sidebands until August 2006.
For the high-resolution (0\farcs23) continuum image, we used the USB data of the full 2 GHz bandwidth
to obtain the highest spatial resolution and sensitivity. 
The CO leakage to the USB should be negligible in the data because the line emission was
heavily resolved out at the resolution.
We did not find a difference except for sensitivity between the 
2 GHz-bandwidth map and a test map made only using the band-edge channels.
We refer to both high and low resolution data  as 860 \micron\ continuum for simplicity
despite their slightly different wavelengths.

Regarding possible contamination of the continuum by weak lines other than CO, 
the large velocity width (\about1100 \kms)
of the galaxy prevented us from noticing such weak lines overlapping with each other.
The upper sideband contained a part of the \HCOplus(4--3) line redshifted to 350.37 GHz, but
its contribution to the continuum flux must be small.
The expected flux of the line is estimated to be 150 Jy \kms\ 
from the galaxy's HCN(4--3) flux of 260 Jy \kms\ \citep{Wiedner02}
and the \HCOplus(1--0)/HCN(1--0) intensity ratio of 0.6 \citep{GraciaCarpio06}.
The continuum flux in the 2 GHz bandwidth is about 1400 Jy \kms. 
Thus the contamination of the USB continuum by the line is expected to be below 10\% 
considering that only a part of the line, velocities larger than 5200 \kms, is in the sideband.
The actual USB spectrum in Fig. \ref{f.spectra} does not convincingly show the line, although
a hint of excess emission may be around 5500 \kms.
We thus disregard this possible line contribution.
The upper sideband should also contain an edge of the HCN(4--3) line (velocities below 5000 \kms),
but it can hardly affect the continuum flux.

\subsubsection{Continuum Subtraction}
The continuum determined from the LSB line-free channels was subtracted from the CO(3--2) line
data in the same sideband. The subtraction was made in the visibilities.
The need for this subtraction is not guaranteed if there is strong CO emission at the continuum location,
because the continuum source may be mostly covered by molecular gas that is optically thick in CO.
An unnecessary subtraction of the continuum would produce an artificial depression in the CO maps 
at the location of each continuum peak.
In reality, our continuum-subtracted CO maps did not show a clear depression. 
Hence we use the continuum-subtracted CO data.
The lack of clear CO depression at the continuum positions in the 
continuum-subtracted images is indicative of less than complete
coverage of the continuum source by the CO-emitting gas. 
Such partial coverage is consistent with the nuclear disk model.
The gas coverage at each position  would be
limited to a small velocity range in the model, making the velocity-averaged coverage
insignificant in the integrated intensity maps.

\subsection{Comparison with Single-dish Observations}
\label{s.red.comparison}
The CO(3--2) flux  of Arp 220 is $(2.2 \pm 0.3) \times10^3$ Jy \kms\ and $(3.2 \pm 0.5) \times10^3$ Jy \kms\
in the central 3\arcsec\ and 5 \arcsec, respectively, 
in our 0\farcs5 resolution data (Fig. \ref{f.spectra}). The errors are mostly due to flux calibration uncertainty. 
The line flux in the JCMT single-dish observations by \citet{Wiedner02} is $(2.98 \pm 0.6) \times 10^{3}$ Jy \kms\
for the central 14\arcsec\ and $(3.7 \pm 0.7) \times10^{3}$ Jy \kms\ for the central 22\arcsec\  \footnote{
	Other single-dish CO(3--2) fluxes in the literature are in the range of 
	$(3.0 \pm 1.5)\times 10^{3}$ Jy \kms\
	\citep{Gerin98,Mauersberger99,Yao03,Narayanan05}. 
	 The scatter among the observations may be due to, 
	 in addition to the difficulty in submillimeter calibration,
	 the broad line (\about1100 \kms\ = 1.3 GHz) that is too wide for some spectrometers.
	\citet{Wiedner02} had a sufficiently large bandwidth with 1.86 GHz.
}.
Thus the SMA observations detected most of the line emission in the
central few kpc of the merger though some of the extended emission 
was likely resolved out. 
The SMA line profile in Fig. \ref{f.spectra} matches
the JCMT spectrum taken at the galaxy center, both being double-peaked and having
a higher intensity in the blueshifted peak as in CO(2--1).
The CO(3--2) flux detected in our 0\farcs33 resolution data is $(1.6 \pm 0.2) \times10^3$ Jy \kms\ 
As expected the high resolution map resolved out more flux.

The 345 GHz continuum in the central 3\arcsec\ of Arp 220 is $0.78 \pm 0.12$ Jy  and $0.72 \pm 0.11$ Jy 
in our 0\farcs5 and 0\farcs23 resolution images, respectively. 
Single-dish observations of the galaxy at 850 \micron\  give a flux of about 0.8 Jy with a typical error
of 20\% 
(0.75 Jy, \citealt{Seiffert07}; 0.74 Jy, \citealt{Klaas01}; 0.83 Jy, \citealt{Dunne00}; 0.79 Jy, \citealt{Lisenfeld00}; 
0.83 Jy, \citealt{Rigopoulou96};  an outlier  0.46 Jy  in \citealt{Anton04}).
Contributions of the CO(3--2) line to these bolometer measurements
are not always given, but was estimated to be 11\% in the study by \citet{Klaas01}.
This flux comparison indicates that we detected
almost all of the continuum emission with the SMA.

\section{CO(3--2) line in Arp 220}
\label{s.result.co}

Figures \ref{f.low} and \ref{f.high} show our low (0\farcs5) and high (0\farcs33) resolution
images of the CO(3--2) emission, respectively. 
The two nuclei seen in continuum are marked with plus signs.
Parameters derived from the data are in Tables \ref{t.datasets} and \ref{t.positions}.

\subsection{Spatial Distribution}
\label{s.result.co.distribution}
The CO(3--2) emission peaks around each of the two nuclei (Figs. \ref{f.low}a and \ref{f.high}a)
and has an extended component encompassing the binary nucleus.
This distribution agrees with those found in CO(1--0) and CO(2--1) through
previous interferometric observations at 1\arcsec--0\farcs3 resolutions 
(\S \ref{s.introduction}).
Following \citest{Sakamoto99},
we call the CO peaks associated with the nuclei ``nuclear disks'' and the extended
emission an ``outer disk''.

Both nuclear disks appear resolved in our 0\farcs33 resolution CO map
with elongation in the north-south direction.
The western nuclear disk appears to have twin peaks.
The eastern nucleus shows a lopsided CO distribution with an extension to the north. 
This elongation agrees with that seen in near-IR, where HST images \citep{Scoville98} show 
two components `NE' and `SE' in the east side of the merger. 
These complex distributions of CO emission suggest that the nuclear disks have non-uniform 
and non-axisymmetric gas properties.
The observed distribution of CO(3--2) emission likely reflects not only gas distribution
but also variation of gas temperature and line excitation, missing flux, 
and the blending with the outer disk.

\subsection{Velocity Field}
\label{s.result.co.velocity}
The velocity field of molecular gas in the merger shows
an overall northeast-to-southwest shift (with NE redshifted) in the outer disk 
and rapid velocity shifts across each nucleus.
The apparent velocity gradients across the two nuclei are not aligned (Figs. \ref{f.low}b and \ref{f.high}b).
The east nuclear disk has a NE-SW gradient similar to that of the outer disk while
the west nuclear disk has a west-to-east gradient with western side being redshifted.
The observed velocity structure is consistent with those found in previous observations
and lead to the model of two counter-rotating disks 
(CO, \citest{Sakamoto99}; 
\ion{H}{1} absorption, \citest{Mundell01};
near-IR CO and \HH, \citealt{Genzel01};
H53$\alpha$ \citealt{Rodriguez05}). 
The steep velocity gradient across the eastern nucleus is also seen in OH maser emission \citep{Rovilos03}.
The overall rotation of the outer disk is consistent with that in previous lower-resolution 
observations of molecular gas (\citest{Scoville97},  \citealt{Downes98}) and ionized gas
\citep{Arribas01}.

The width of the CO line is clearly peaked at the two nuclei (Fig. \ref{f.low}c) as expected from the
steep velocity gradient across each nucleus.
The line profiles at the nuclei (Fig. \ref{f.nucleus.spectra}) are either flat-topped or double-peaked.
This was first noticed in CO(2--1) by \citest{Scoville97} and led them to infer the presence of a 
rotating disk around each nucleus. 
The intensity dip near the line center may be also due to absorption by cooler CO near the
surface \citesp{Downes07}.
The line width at 20\% of the maximum intensity in the profiles is 540 \kms\ and 600 \kms\ at the east and
west nucleus, respectively, in our 0\farcs5 resolution data.

The position-velocity (PV) diagrams across each nucleus confirm the large line widths around,
 and large velocity shifts across, the nuclei (Fig. \ref{f.pv.nucleus}).
The line velocity changes by several 100 \kms\ within a few 100 pc from one side of a nucleus 
to the other side.
In the innermost part of the PV diagrams where the apparent velocity gradient is nearly linear, 
we measured a gradient of  2 \kmsperpc\ across the E nucleus 
and 6 \kmsperpc\ across the W nucleus. 
The PV diagram of the west nucleus shows a very large velocity shift in a very small region
around the center, and hence the rotation curve there may not be close to linear.
\citest{Downes07} suggested that the velocity is larger closer to the center 
in the west nucleus in their 0\farcs3 CO(2--1) data. 
Although a hint of a similar trend is seem in our PV diagram (of 0\farcs28 resolution along the cut)
in its redshifted side (offset = 0\asec--0\farcs3;  5350--5550 \kms), 
we conservatively use only the velocity shift
of 500 \kms\ in 0\farcs22 = 80 pc without specifying the shape of the rotation curve.
The (apparent) gradients that we obtained are consistent with those in \citest{Sakamoto99}
but are larger than the \about0.3--1 \kmsperpc\ in previous studies
at similar spatial resolutions (\citest{Mundell01}; \citealt{Rovilos03,Rodriguez05}).
Those lower gradients are probably because they were measured in the mean velocity maps,
in which gradients are falsely reduced through beam smearing. 
Even our own mean-velocity map (Fig. \ref{f.high} b) gives the apparent gradient 
of about 1 \kmsperpc\ for both nuclei, in contradiction with the \about500 \kms\ line widths 
within 0\farcs5 (=180 pc; Fig. \ref{f.nucleus.spectra}).
Also possible, though less likely, is that CO and all the other lines in previous studies 
(\ion{H}{1}, OH, and H53$\alpha$) trace different parts of
the disks.

\subsection{Brightness Temperature, Column Density}
\label{s.result.co.Tb_sigma}
The peak brightness temperature of the CO line is about 50 K.
The highest temperatures are seen in the vicinity of each nucleus in the high-resolution data 
(Fig. \ref{f.high}c).
The peak \Tb\  is slightly lower across each nuclear disk along the isovelocity contours, seen
more clearly in the 0\farcs33 data though also visible in the 0\farcs5 map.
These temperature troughs are most likely due to a lowered filling factor of the molecular gas
in each velocity channel in the region of largest velocity gradient rather than due to an actual
decline of the physical gas temperature, 
because the directions of the isovelocity contours are determined by our viewing angle.
Thus the data suggest that each nuclear disk has a gas temperature of \about50 \KRJ\
when averaged over 120 pc (0\farcs33) without a correction for an area filling factor.
The apparent high-temperature region is larger around the west nucleus and has a radius
of about 100 pc at the 40 \KRJ\ contour. 
The gas kinetic temperature may be higher if the beam filling factor is less than unity 
or the line is not optically thick.
The gas in the outer disk shows lower brightness temperatures than in the nuclear disks, 
but it is still mostly higher than 10 \KRJ\ in the central kpc.
The temperature gradient is probably due to lower filling factor and lower physical temperature 
in the outskirts.

Fig. \ref{f.nucleus.spectra} shows that the CO line profile and brightness temperature are similar 
in the J=3--2 and J=2--1 transitions at each nucleus.
To more accurately measure the line ratio, we made matching datasets having the same minimum 
baseline (41 k$\lambda$) and resolution (0\farcs57) from our SMA observations and the OVRO
observations of \citest{Sakamoto99}.
The ratio of integrated brightness temperatures, $R_{(3-2)/(2-1)}$, is 1.1 and 1.0 at the east
and west nucleus, respectively, with an 18\% error from flux calibration uncertainties.
The ratio of unity is expected for optically thick emission from warm and thermalized CO.

The peak CO(3--2) integrated intensity in our data is as large as \about$1\times10^{4}$ \KRJ\ \kms\ at
the two nuclear disks.
The column density of \HH\ gas is
$\Sigma_{\rm H_2} = 3\times 10^{24}\, m_{\rm H_{2}}\, \persquarecm 
= 5\times 10^4$ \Msol\ \persquarepc\ for a Galactic conversion factor 
of $N_{\rm H_2}/I_{\rm CO}=3\times 10^{20}$ \unitofX, the applicability of which to ULIRGs is not guaranteed.
A lower limit to the mass surface density of \HH\ can be obtained assuming that
the CO emission is optically thin.
For an abundance of $[{\rm CO}/\HH]=10^{-4}$ and 50 K LTE excitation, the
limit is 
$\Sigma_{\rm H_{2}} \geq 5\times 10^{22}\, m_{\rm H_{2}}\, \persquarecm = 7\times 10^2$ \Msol\ \persquarepc.
The excitation temperature of 50 K is a lower limit set by the observed
brightness temperature;
the higher the excitation temperature, the higher the lower limit of $\Sigma_{\rm H_{2}}$.
We set a more stringent lower limit later using the CO equivalent width 
(\S \ref{s.nuclei.ISMproperties_fromWeq}).
The gas surface density including helium can be obtained
by multiplying $\Sigma_{\rm H_{2}}$  by 1.36.

\section{860 \micron\ continuum in Arp 220}
\label{s.result.continuum}
Figs.  \ref{f.low} and \ref{f.high} show our 860 \micron\ continuum maps of Arp 220 
at low (0\farcs5) and high (0\farcs23) resolutions, respectively.
The  continuum has two peaks at the two nuclei and is much more compact than
the CO line emission.

\subsection{Positions}
\label{s.result.continuum.positions}
The positions of the nuclei in the continuum, listed in Table \ref{t.positions},
were measured in the two data sets through Gaussian fitting
and were averaged for each nucleus. 
The positions from the low- and high-resolution maps agree to within 0\farcs02 for the west
nucleus and to 0\farcs08 for the weaker (and likely more extended) east nucleus.
The average positions agree within 0\farcs05 with the 1.3 mm positions in \citest{Sakamoto99}.
The two nuclei are separated by 0\farcs98 at a position angle of 101\arcdeg.

The 860 \micron\ as well as 1.3 mm continuum positions agree well with centimeter VLBI observations 
in terms of the relative position between the two nuclei, 
but show a slight (\about0\farcs1) offset most likely due to astrometric errors.
The median coordinates of the cluster of supernova-related compact ($< 1$ pc = 3 mas) radio sources 
around each nucleus  are
$\alpha$ = 15$^{\rm h}$34$^{\rm m}$57\fs2224,  $\delta$  = \plus23\arcdeg  30\arcmin 11\farcs492    
for the 31 sources around W and
$\alpha$ = 15$^{\rm h}$34$^{\rm m}$57\fs2910,  $\delta$  = \plus23\arcdeg  30\arcmin 11\farcs308    
for the 21 sources around E.
These are calculated from the source lists in \citet{Parra07} and \citet{Lonsdale06} 
and are in the coordinate system used in \citet{Parra07} who 
reported that they fixed an error in previous observations.
The offset of E from W, ($\Delta\alpha, \Delta\delta$) = (0\farcs94, \minus0\farcs18), 
agrees well with that from the SMA. 
However, both nuclei show an offset of $0\farcs10 \pm 0\farcs01$ in almost the same direction 
between the SMA and the centimeter VLBI positions. 
The offset is $0\farcs11 \pm 0\farcs02$ between the \citest{Sakamoto99} and the VLBI positions.
Attributing these offsets to astrometric errors, 
in which ours may be significant (see \S \ref{s.red.seeing}),
we shift maps to cancel the offset when we later compare the spatial distribution of the
centimeter sources and submillimeter emission.

\subsection{Spatial Distribution}
\label{s.result.continuum.distribution}
The two nuclei account for about 75\% of the total 860 \micron\ continuum in the central 3\arcsec.
The rest is in the outer disk but its surface brightness is too low for us to convincingly discern its structure.
The flux ratio between the two nuclei is about $1:2$ between E and W.

Each of the two continuum nuclei is partially resolved in our data and has a deconvolved size 
of  \about0\farcs15--0\farcs3 (Table \ref{t.datasets}).
The beam-deconvolved sizes of the nuclei in our high-resolution map 
are larger than the finite deconvolved size of a point source, 0\farcs09,
caused by atmospheric seeing and calibration errors (\S \ref{s.red.seeing}). 
Thus, the submillimeter continuum of each nucleus is intrinsically extended to 
a few tenths of an arcsecond.
The deconvolved size of the west nucleus from our 0\farcs23 map agrees with
the value that \citest{Downes07} obtained from their 0\farcs3-resolution map of 1.3 mm continuum.

The east nucleus has a larger deconvolved size than the west and has a major axis roughly in the
northeast-southwest direction in both of our continuum maps.
Interestingly, an elongation similar in size and direction is seen and noted in centimeter continuum observations (\citealt{Baan87}; \citealt{Norris88}; \citealt{Baan95}; \citest{Mundell01}; \citealt{Rovilos03}),
mid-IR imaging \citesp{Soifer99},
and is also noticeable in the distribution of the radio supernovae and young supernova remnants
\citep{Lonsdale06,Parra07}.
The west nucleus, on the other hand, does not show the east-west elongation seen in
the distribution of the centimeter compact sources.
We discuss this in \S \ref{s.discuss.energy}.

Our imaging results are mostly consistent with the results of the previous single-baseline interferometry 
at the same wavelength by \citet{Wiedner02}.
Analyzing the visibilities from the 164 m JCMT-CSO baseline,
they found that the 342 GHz continuum could be modeled as a binary system with an east-west separation of  \about1\arcsec\ 
and a flux ratio of $1:1.7$,
and that the CO(3--2) line was more extended than the continuum. 
These agree with our observations.
The total continuum flux attributed to the binary nuclei is $0.40 \pm 0.09$ Jy in the JCMT-CSO observations
and $0.55 \pm 0.08$ Jy in our data.
Although the former makes more room for extended emission 
when compared to the total continuum flux density of about 0.8 Jy measured with single-dish telescopes,
the JCMT-CSO visibility data do not hint at such extended continuum.
Namely, the continuum visibilities did not show larger amplitudes when the baseline was most foreshortened 
and had a fringe spacing as large as 5\arcsec\ \citep[Fig. 6 of][]{Wiedner02}.
This was in contrast to the CO visibilities that clearly showed larger amplitudes on
foreshortened baselines to suggest extended emission.
Thus the JCMT-CSO observations and our SMA observations agree that
the CO emission has a significant extended component while the continuum lacks it.

\subsection{Brightness Temperature and Spectral Index}
\label{s.result.continuum.Tb_spindex}
The two nuclei have high brightness temperatures in the continuum, indicating the presence of warm dust.
The  continuum \Tb\  calculated using the Planck function
is 27 K and 54 K for the east and west nuclei, respectively, in the 0\farcs23 resolution map.
The temperatures are higher when the sources are deconvolved.
A marginally resolved source such as these nuclei can be deconvolved equally well 
as a uniform-brightness disk or a Gaussian,
and the disk diameter is about 1.6 times the Gaussian FWHM for the same source
(Appendix \ref{a.disk_deconvolution}).
The west nucleus has a FWHM size of \about0\farcs15 = 53 pc and
a peak brightness temperature of  $1.6\times10^2$ K  in the Gaussian model,
and a diameter of 0\farcs23 = 85 pc and a temperature of 91 K in the disk model.
The east nucleus has a peak temperature of 52 K and 29 K for Gaussian and disk models, respectively.
The actual dust temperature must be even higher than these if the emission is optically thin or
blurred by atmospheric seeing.

The spectral index of the continuum emission between 230 GHz (1.3 mm) and 345 GHz (860 \micron) is 
estimated to be $\alpha_{230-345}({\rm E}) =2.6 \pm 0.6$ 
and $\alpha_{230-345}({\rm W})=2.7\pm0.5$ 
for the east and west nucleus, respectively, by 
comparing our SMA observations with the OVRO 1.3 mm data of  \citest{Sakamoto99}.
The index $\alpha$, defined as $S_\nu \propto \nu^\alpha$, is calculated 
from peak intensities of each nucleus in $0\farcs56 \times 0\farcs51$ resolution images at the two frequencies.
The SMA map used for this was matched to the OVRO data by eliminating the short
baselines ($\leq$ 40 k$\lambda$) and tapering long baselines.
The errors of the spectral indices include those from the noise in the maps and the flux
calibration uncertainties.
The indices for dust emission only are larger by 0.1 than the above values.
These are calculated by subtracting the non-dust emission,
which we estimate from radio data in the literature 
to be $10\pm 5$ ($5\pm 5$) mJy at both 230 and 345 GHz
for the west (east) nucleus.

The spectral indices we obtained at the nuclei are larger than that of a blackbody ($\alpha \leq 2$) 
and smaller than the index for the entire merger at these frequencies 
($\alpha \approx 3.5$; \citealt{Scoville91}; \citealt{Downes98}).
The latter index is usually attributed to optically thin emission from dust 
with an emissivity index of $\beta \sim$  1--2 (for $\epsilon_\nu \propto \nu^{\beta}$) and one or
more temperature components \citep{Eales89,Rigopoulou96,Lisenfeld00,Dunne00,Dunne01,Klaas01,Blain03}.
Our smaller spectral indices toward the nuclei are suggestive of
moderate optical depths (i.e., $\log (\tau_{\rm 1\, mm}) \sim 0$) toward the nuclei.
We attempt a more quantitative assessment below.

\subsection{Optical Depth and Dust Temperature}
\label{s.result.continuum.tau_Td}

\subsubsection{West Nucleus}
We constrain the optical depth and dust temperature of the west nucleus using the dust sublimation
temperature, the spectral index around 1 mm, and the comparison of our submillimeter data
with mid-IR observations.
First, a lower limit to the dust optical depth can be set from the sublimation temperature of dust,
\about2000 K \citep{Salpeter77, Phinney89}.
For Arp 220 W, 
the maximum dust temperature and the lower of the deconvolved brightness temperatures 
at the nucleus 
limit the temperature of 860 \micron\ emitting dust to 90 K $\leq  T_{860} \leq$ 2000 K
and 
dust optical depth at 860 \micron\ to $\tau_{860} \geq 0.05$.

The next constraint on the optical depth, that  $\tau_{860}$ is of the order of 1 or less,
is set from the comparison of 1300 \micron\  and 860 \micron\  data, 
as we saw in the previous section.
The ratio of Rayleigh-Jeans brightness temperatures between two wavelengths $\lambda_1$ and $\lambda_2$
is expressed as
\begin{equation}
	\label{eq.R_mm}
	R_{\lambda_2/\lambda_1}
	\simeq
	\frac{1-\exp(-\tau_2)}{1-\exp[-(\lambda_2/\lambda_1)^{\beta}\tau_2]},	
\end{equation}
where $\tau_2$ is the optical depth at $\lambda_2$. 
Any emission not originating from dust has to be subtracted before calculating the ratio.
This formula using the Rayleigh-Jeans approximation is accurate to 3\% 
for the minimum dust temperature of 90 K and the two wavelengths we consider.
The ratio $R_{\lambda_2/\lambda_1}$ is 
1 if the emission is optically thick in both wavelengths, 
$(\lambda_1/\lambda_2)^{\beta}$ if both are thin, 
and in-between for moderate opacities.
The 0\farcs5-resolution OVRO and SMA data of the west nucleus 
give $R_{860/1300}({\rm dust})=1.40 \pm 0.30$. 
The ratio is also calculated from our 0\farcs23 data and the 0\farcs3-resolution data of \citest{Downes07} 
to be $R_{860/1300}({\rm dust})=1.57 \pm 0.32$ for the compact west nucleus; 
here we arbitrarily used 10\% for the uncertainty of the 1300 \micron\ flux including
any error caused by different \uv\ coverage.
The limits on the opacities from these ratios and Equation (\ref{eq.R_mm}) 
are summarized in Table \ref{t.opacity}.
For example, the former ratio and $\beta=2$ give $\tau_{860}=2.3$ as the most likely value
and $\tau_{860}=$(1.2--5.3) for the range that 
corresponds to the $\pm 1\sigma$ error of  $R_{860/1300}$.
Extending this analysis to longer wavelengths is not easy because one has less dust emission and
more free-free and synchrotron emission in the continuum.

Further constraints can be set on the optical depth and dust temperature 
by combining our submillimeter data with the 25 \micron\ imaging photometry 
as \citest{Soifer99} demonstrated.
The model has a bright hot nucleus behind a cool shallow absorber whose absorption is
significant only in mid-IR and emission is insignificant in both wavelengths.
Our analysis deviates from that of \citest{Soifer99} (and \citest{Downes07}) in that the nucleus is
allowed to have different apparent sizes between 25 \micron\ and 860 \micron,
to be consistent with the latest observations. 
The ratio of 860 \micron\ and 25 \micron\ flux densities from the dust can be written as 
\begin{equation}
	\label{eq.R_mir}
	\frac{S_{25}}{S_{860}} 
	\approx
	\frac{\exp(-\tau_{25,{\rm fg}})}{1 - \exp(-\tau_{860})}
	\frac{B(25, T_{25})}{B(860, T_{860})} 
	\frac{\Omega_{25}}{\Omega_{860}},
\end{equation}
where $S_{\lambda}$ and $\Omega_{\lambda}$ are the observed flux density of 
and the solid angle subtended
by the nucleus, respectively, at wavelength $\lambda$ \micron, $B$ is the Planck function, 
$T_{25}$ is the dust temperature at the 25 \micron\ photosphere of the nucleus,
 and
$\tau_{25,{\rm fg}}$ is the foreground opacity that is likely to be \about1--2 
(\citest{Soifer99}; see also \citealt{Haas01}; \citealt{Spoon04}).
The nucleus is almost certainly optically thick at 25 \micron\ because  
$\tau_{860} \geq 0.05$ and $\tau_{25} \simeq \tau_{860}(860/25)^\beta$ with $\beta$ in the range of 1--2.
The dust in the compact core seen at 860 \micron\
is  assumed to be at a temperature of about $T_{860}$.
Our SMA data and the 25 \micron\ data of \citest{Soifer99} give
$S_{25}/S_{860}({\rm dust}) =21$ and $\Omega_{25}/\Omega_{860}=$3.0--7.3, where the
latter range corresponds to the possible FWHM size of the 25 \micron\ emission of 
$\theta_{25}=$0\farcs25--0\farcs39 \citesp{Soifer99}.
Further assuming that the dust at the 25 \micron\ photosphere of the nucleus
is heated entirely by the more compact 860 \micron-emitting core,
we have $T_{25}/T_{860}=$ 0.76 and 0.61 for $\theta_{25}=0\farcs25$ and 0\farcs39, respectively, 
since the temperature would be inversely proportional to the square root of radius in this circumstance.
These inputs to Eq. (\ref{eq.R_mir}) give an upper limit to the dust temperature in the nucleus.
Namely, $T_{860} \leq 1.4\times 10^2$ K and $1.5\times 10^2$ K for 
$\theta_{25}=$0\farcs25 and 0\farcs39, respectively, if $\tau_{25,{\rm fg}}$ is 1.
Limits for $\tau_{25,{\rm fg}}=2$ are in Table \ref{t.temperature}.
The upper limits would be lower if there are additional heating sources
between the 860 \micron-emitting core and the 25 \micron\ photosphere.
For completeness, Table \ref{t.temperature} also has an upper limit to the dust temperature  
for the case in which the 25 \micron\ and 860 \micron\ emission originates
from exactly the same surface and hence $\Omega_{25}=\Omega_{860}$ and $T_{25} = T_{860}$.
Because the upper limit temperature to the 860 \micron-emitting dust does not exceed $1.8\times 10^2$ K
for the plausible combinations of parameters, we obtain a lower limit to the 860 \micron\ opacity of
the west nucleus, $\tau_{860} \ge 0.7$, from the 90 K lower limit to 860 \micron\ brightness temperature.

The constraints set above are summarized as 90 K $\leq T_{860} \leq$ 180 K and
$0.7 \leq \tau_{860} \leq 5$ for the submillimeter-emitting dust in the west nucleus.
Given our crude modeling, 
in which a single opacity is assigned to an object that is unlikely to be a uniform slab, 
the latter should be taken more loosely to mean that  $\tau_{860}$ of the dust that contributes
most to the submillimeter emission is of the order of 1.
We call attention to the fact that the temperature will be on the lower 
side of the 90--180 K range if there are additional heating sources such as massive stars
between the bright 0\farcs15 core seen at 860 \micron\ and the  0\farcs25--0\farcs39 photosphere
at 25 \micron. The 860-\micron\ opacity will then be on the higher side.
The moderate optical depth in submillimeter toward the nucleus (about 1 when converted to $\tau_{860}$)
has been suggested by several authors (\citest{Soifer99}; \citealt{Gonzalez04}; \citest{Downes07}).

\subsubsection{East Nucleus}
The east nucleus has less constraining data but its 860 \micron\ opacity is
also likely to be on the order of 1. 
Here, the nucleus has $R_{860/1300}({\rm dust})=1.34 \pm 0.36$ at 0\farcs5 resolution.
The formal solution of Eq. (\ref{eq.R_mm}) is $\tau_{860} = 0.8$ and 2.8 for $\beta=1$ and 2, respectively.
The east nucleus has  $S_{25}/S_{860}({\rm dust}) =16$.
It is unresolved at 25 \micron\ with 0\farcs62 resolution, but shows elongation or band-dependent
positional shift in shorter infrared wavelengths (\citest{Soifer99}, \citealt{Scoville98}).
The foreground extinction to the east nucleus is 50\% larger than to the west nucleus \citesp{Soifer99}.
This may be the cause of the lower flux ratio and color temperature than for the west nucleus.
If so, the actual dust temperature does not need to be lower than in W.
For the simplest assumption of $\Omega_{25}=\Omega_{860}$ and $T_{25} = T_{860}$,
we obtain temperature upper limits of (0.9, 1.2, 1.6)$\times 10^{2}$ K 
for $\tau_{25,{\rm fg}}=0$, 1, and 2, respectively.

\subsubsection{Constraint from X-ray Observations}
The optical depth of $\tau_{860} \sim 1$ toward the nuclei corresponds to
a column density of $N_{\rm H_2} \sim 4\times10^{25}$ \persquarecm, 
and a hydrogen mass surface density of \about$6\times10^{5}$ \Msol\ \persquarepc\
if the Galactic dust emission law of \citet{Hildebrand83} applies.
(Our expression of column density in $N_{\rm H_2}$ is for convenience and not
specifying the phase of the hydrogen.)
The column density of $10^{25}$ \persquarecm\ can block the X-rays from
an AGN that is energetically significant to Arp 220 and hence could explain
the non-detection of such X-rays \citep{Rieke88,Dermer97,Iwasawa01}.  
This suggests another possible constraint on $\tau_{860}$. 
Namely, if $\tau_{25,{\rm fg}}$ is larger than the values we used 
then $T_{860}$ could be higher and hence $\tau_{860}$ lower.
The bolometric luminosity of the brighter west nucleus would then exceed $10^{12}$ \Lsol.
For example, it would be at least $10^{13.0}$ \Lsol\ and  $10^{13.7}$ \Lsol\ 
for $\tau_{860}=0.5$ and 0.3, respectively (\S \ref{s.result.continuum.luminosity}).
An AGN would be needed to explain, among other things, a luminosity to mass ratio that is too high for a starburst 
(\S \ref{s.nulcei.energy_source.L/M_and_Sigma_L}).
A luminous AGN and too low $\tau_{860}$ would however contradict
the X-ray constraint mentioned above, provided that the dust covering the AGN 
emits the bulk of the submillimeter emission.  
Thus the submillimeter opacity is not likely to be much smaller than 1,
although a precise lower limit is difficult to set this way with poorly known 
dust properties, dust to gas ratio, and column density distribution in the nucleus.

\subsection{Luminosity}
\label{s.result.continuum.luminosity}
The bolometric luminosity of each nucleus can be calculated from its temperature and size.
The luminosity of a blackbody sphere of diameter $d$ and temperature $T_s$ is
\begin{equation}
	L_{s}
	=
	\pi d^2 \sigma T_s^4,
\end{equation}
where $\sigma$ is the Stefan-Boltzmann constant.
For comparison, the luminosity of a blackbody whose surface brightness profile on the sky 
is a Gaussian of FWHM size $\theta_{\rm FWHM}$ and peak temperature $T_g$ is
\begin{equation}
	\label{eq.gaussian_luminosity}
	L_{g}
	\approx
	\frac{\pi \theta_{\rm FWHM}^2}{4\ln 2} \sigma T_g^4
\end{equation}
if the power we receive is about the average of direction-dependent power from the source.
As noted earlier,
a marginally resolved source can be modeled both as a uniform-brightness disk and a Gaussian,
and the disk diameter and the Gaussian FWHM for the same source have the relation
$d \approx 1.6\,\theta_{\rm FWHM}$ (Appendix \ref{a.disk_deconvolution}).
The (peak) temperatures in these two models have the following relation
\begin{eqnarray}
	T_s &=& \frac{1}{\ln 2} \left( \frac{\theta_{\rm FWHM}}{d} \right)^2 T_g
	\\ 
	&\approx& 0.56 \, T_g	\qquad \mbox{(for $d=1.6\,\theta_{\rm FWHM}$)}.
\end{eqnarray}
The bolometric luminosities have the relation
\begin{equation}
	L_{s} \approx 0.72 \, L_g
\end{equation}
for the same submillimeter flux density, or for the same flux density at any frequency where 
the Rayleigh-Jeans approximation applies.
In addition, a thin disk of the same diameter $d$ has a total luminosity 
$L_{d} \approx 0.5 L_s$ if we observe it from pole-on, and $L_{d} > L_s$ when edge-on.
Note that the luminosity estimated from the submillimeter flux density ($S_{\nu}$)
strongly depends on the size ($R$) and temperature ($T$) of the source of any profile as
\begin{equation}
	L 
	\propto
	R^2 T^4	
	\propto
	S_\nu^4 R^{-6}
	\propto
	S_\nu T^{3}.
\end{equation}
Hence, great care is needed in size and temperature determination for the luminosity estimate.

Our 860 \micron\ data 
give $3.1 \times 10^{11}$ \Lsol\ and $2.2 \times 10^{11}$ \Lsol\ for
the total luminosity of the west nucleus in the Gaussian and sphere models, respectively, under
the blackbody approximation.
The luminosity would be larger than that for the sphere if the 860 \micron\ source was a disk and nearly edge-on,
which has been suggested for the western nuclear disk encompassing the
submillimeter core (\citealt{Scoville98}, \citest{Sakamoto99}).
For comparison, the blackbody luminosity is  (3--4) $\times 10^{10}$ \Lsol\
for the 0\farcs3-resolution 1.3 mm data of \citest{Downes07}.

These luminosities are lower limits to the true bolometric luminosity of the nucleus for two reasons.
First, the size used to estimate the luminosity may be larger than the true size because
a point source would have a finite size of 0\farcs09 in our data due to atmospheric seeing 
and various calibration errors (\S \ref{s.red.seeing}).
Removal of this would make the source FWHM, peak temperature, and the total luminosity
0\farcs11 = 40 pc, $2.7\times10^2$ K, and $1.4 \times 10^{12} \Lsol$, respectively, for the Gaussian model.
We warn however that these numbers, in particular the luminosity, have large uncertainties and that
they should be taken only as possible values until supported by higher-resolution observations.
The second but more important reason why the (2--3)$\times 10^{11}$ \Lsol\ is a lower limit is that
our 860 \micron\ brightness temperatures are lower limits to the true temperature of the dust
because of the moderate opacity of the dust emission.
The dust temperature would be 1.6 times higher than the observed brightness temperature
if $\tau_{860}$ is 1, 
and that would make the bolometric luminosity \about6 times larger to $L \sim 10^{12} \Lsol$.
The use of the Stefan-Boltzmann law in this calculation is valid 
because most of the luminosity is emitted in mid- to far-IR where the dust is 
opaque\footnote{
	The luminosity of a gray body with a frequency-dependent opacity 
	of $\tau(\nu)=(\nu/\nu_1)^\beta$ ($\beta=$1--2) 
	is more than 90\% of the blackbody luminosity for the same temperature 
	if $h\nu_1/kT \le 1$, where $\nu_1$ is the frequency at which $\tau$ is 1.
}.
Although the opacity correction increases the luminosity by only 3\% if $\tau_{860}=5$,
it is almost certain, as can be inferred from our CO maps, that the dust around the nucleus
has a lower-column density periphery that is transparent at 860 \micron\ but is opaque
at shorter wavelengths.
This luminosity from outside of the bright 860 \micron\ core but within the nuclear disk
is probably going to increase the luminosity of the nucleus to the level expected in the
current model of luminosity distribution in the merger system (\S \ref{s.introduction}).
The extra luminosity of the periphery would be either reemission of  the radiation from the hot core 
or the luminosity of the starburst in the nuclear disk.

The east nucleus has a (minimum) luminosity of \about$5\times 10^{9}$ \Lsol\ before
any seeing and opacity corrections.  
The luminosity of the outer disk is poorly constrained by our continuum observations. 
Our line data, however, suggest that it may be as large as a few $10^{11} \Lsol$ 
as in the currently prevailing model described in \S \ref{s.introduction}.
The outer gas disk of \about1 kpc diameter and
several $10^{3}$ \Msol\ \persquarepc\ surface density \citesp{Scoville97} is
opaque shortward of \about50 \micron\ and will
have a total luminosity of a few $10^{11}$ \Lsol\ if it is at 40 K.
The high peak brightness temperatures of the CO(3--2) line, 
$\gtrsim 30$ K, seen in the central 2\arcsec\ (0.7 kpc) of the merger hint at such a warm ISM
in the outer disk.
In addition to the luminous nuclei, the outer disk has its own young massive clusters for heating
\citep[e.g.,][]{Scoville98,Wilson06}.

\section{Properties of the Nuclei}
\label{s.discuss}
\subsection{Mass and Dynamical Properties}
\label{s.nuclei.dynamical}
The rotating disk around each nucleus is the currently accepted scheme,
although there were alternative models attributing the velocity structure to velocity dispersion
or non-circular motion \citep[e.g.,][]{Downes98,Eckart01}.
For the west nucleus, for example, HST/NICMOS images imply an inclined disk
of absorption (i.e., gas and dust) elongated in the east-west direction \citep{Scoville98}.
The major axis and the extent of the disk agree with those of the observed gas velocity gradient.
Radio supernovae and their remnants in the west nucleus are in a 160 pc by 70 pc region
elongated in the east-west direction \citep{Smith98,Lonsdale06,Parra07}, also
consistent with the inclined and starbursting nuclear disk elongated in the same direction.
\citest{Downes07} also adopt a rotating disk.

The dynamical mass of each nucleus is \about$10^{9}$ \Msol\ within \about100 pc radius
(see \S\ref{s.result.co.velocity} and Table \ref{t.datasets}) as was previously derived from CO(2--1)
and near-IR observations (\citest{Sakamoto99}, \citealt{Genzel01}).
The estimate should be accurate to within a factor of a few, but it is difficult to
tell in which direction it is likely to err.
On one hand, the calculated mass underestimates the true mass 
because no inclination correction is applied,
though the correction is likely small for the highly inclined west nuclear disk.
On the other hand, the radius at which the rotational velocity peaks in each PV diagram 
is quite possibly overestimated 
because each nuclear disk is only marginally resolved in space;  
the dynamical mass will be overestimated by this error.
We can not tell which of the two sources of bias is larger, and hence the calculated mass is neither
a lower limit nor an upper limit.
Still, the current data do not support a small mass of \about$2\times10^{8}/\sin^2 i$ \Msol\
within about 100 pc from the west nucleus suggested by \citest{Mundell01}.
The small velocity gradient seen in \ion{H}{1} absorption that lead to this mass estimate is incompatible
with the large CO velocity range seen in our PV diagram (\S \ref{s.result.co.velocity}).
Their model of merger configuration and history are partly based on the observation that
the west nucleus is an order of magnitude lighter than the east nucleus, 
but that disparity is not supported by our data.
If there is any disparity, the west nucleus {\it appears} to be denser than the east nucleus 
as suggested from the larger velocity gradient of the former (see below), 
but that could be due to a lower inclination of the eastern disk.

The mean mass density calculated from the $V/R$ and the following formula,
which does not assume a linearly-rising or any specific rotation curve,
is $2\times10^2$ \Msol\ \percubicpc\ for $r \leq 110$ pc of the east nucleus
and $2\times10^3$ \Msol\ \percubicpc\ for $r \leq 40$ pc of the west nucleus.
\begin{equation}
	\left( \frac{\overline{\rho}}{\Msol\, \percubicpc} \right)
	=
	55
	\left( \frac{V/R}{\kms / \rm pc} \right)^2
\end{equation}
Elliptical galaxies and bulges have similar mass densities at the same scales \citep{Lauer95}
and our Galaxy has $\overline{\rho}\, (r=12\; {\rm pc}) = 4 \times10^3$ \Msol\ \percubicpc\
\citep{Genzel96}.
The mass densities that we derived are lower limits, because
both the underestimate of the rotational velocity $V$ due to the lack of inclination correction
and the overestimate of the radius $R$ due to insufficient spatial resolution
make the ratio $V/R$ and hence the mean mass density smaller.
For the same reason, we obtain  upper limits of 1--3 Myr to 
the dynamical time-scales (= rotation periods) of the nuclear disks from the velocity-radius pairs.

\subsection{ISM Properties Suggested from Low CO(3--2) Equivalent Width}
\label{s.nuclei.ISMproperties_fromWeq}
In addition to the high brightness temperature, 
our observations show that the nuclear
disks have a much lower line-to-continuum ratio than the outer disk in the submillimeter,
as was the case at 1.3 mm \citesp{Sakamoto99}.
Specifically, about 75\% of the total (i.e., single-dish) 860 \micron\ continuum comes from the nuclear disks while 
only about 25\% of the total CO(3--2) emission is from the nuclear disks. 
The rest of the emission is from the outer disk. 
To put it another way, the equivalent width for the CO(3--2) line is 
\about$1\times 10^3$ \kms\ in the nuclear disks,
while that of the outer disk is several times larger, \about$6\times 10^3$ \kms.

The larger equivalent width of the outer disk is in line with the observations of other
IR bright galaxies  
and the small equivalent width (i.e., low line-to-continuum ratio) in the nuclear disks is exceptional.
For example, \citet{Seaquist04} showed that CO(3--2) emission typically comprises
25\% of 850 \micron\ flux
detected within a \about$3\times10^4$ \kms\ bandwidth 
at the central kiloparsecs of relatively IR-luminous galaxies
($\log (L_{\rm IR}/\Lsol) \approx$ 10--12) in the SLUGS survey \citep{Dunne00,Yao03}.
The fraction of 25\% means a CO(3--2) equivalent width of $1\times 10^4$ \kms.
Similarly, the equivalent widths in the central 15\arcsec\ (250 pc) of
the starburst galaxies NGC 253 and M82 are
$(1.5\pm 0.3) \times 10^4$ \kms\ and $(1.1\pm 0.2) \times 10^4$ \kms, respectively,
according to the 345 GHz observations of these galaxies \citep{Alton99,Israel95,Tilanus91} 
and the prescription by \citet{Seaquist04} to remove the CO(3--2) contribution from the 850 \micron\ bolometer flux.

A simple model tells us that the low equivalent width is basically due to either
high temperature or high column density per velocity.
Figure \ref{f.lte} shows LTE calculations for CO(3--2) optical depth, 
excess brightness temperature, and equivalent width. 
The model assumes that gas and dust in the model region have the same uniform temperature, 
gas-to-dust mass ratio of 100, 
CO abundance of [CO/\HH]=$10^{-4}$, 
and
dust mass opacity coefficient at the CO(3--2) frequency of $\kappa= 1$ \unitofkappa,
and also assumes  
that the dust emission is optically thin. 
Appendix \ref{a.equiv-width} gives model formulae and the reason why the moderate dust optical depth
of the nuclear disks does not affect the derived gas column density much.
Figure 7 shows that there are two possible sets of  gas conditions for each set of equivalent width
and brightness temperature.
One has optically thin CO emission from warmer gas with lower column density per velocity,
and the other has optically thick CO emission from cooler gas with higher column density per velocity.

The optically thin solution is unlikely for the nuclear region of Arp 220
because it needs a high temperature of about  800 K for the nuclear disks.
The required temperature would be even higher 
for plausible non-LTE cases in which levels much higher than J=3 are less populated
than in LTE because of their high critical densities.
Such high temperatures are incompatible with the upper limit set on the dust temperature
in \S \ref{s.result.continuum.tau_Td}.
Moreover, the CO(3--2) to (2--1) intensity ratio $R_{(3-2)/(2-1)}$ is close to 1 at the two nuclei, 
consistent with optically thick emission.
The ratio would be $\frac{9}{4}$ if both lines were optically thin. 
Accepting that the CO(3--2) emission from the nuclear disks is optically thick,
the bulk of the CO emission from the outer disk must also be optically thick because
the intensity ratios between low-J CO lines are about 1 for the entire galaxy
\citep{Mauersberger99,Wiedner02}.
Although the optically-thin solution of 800 K gas is unlikely,
it provides a lower limit to the mean gas surface density in the nuclear disks, 
$\Sigma_{\rm H_{2}} > 4\times 10^{23}\, m_{\rm H_{2}}\, \persquarecm = 6\times 10^3$ \Msol\ \persquarepc,
because this solution has less gas surface density than the optically thick solution.
This is a more stringent lower limit to the column density than the one in \S \ref{s.result.co.Tb_sigma},
thanks to the additional constraint from equivalent width.

In the preferred case of optically thick CO emission, the low equivalent width is due to
a high column density per velocity width, 
$N_{\rm CO}/\Delta V \approx 10^{18.5}$ \persquarecm\ \perkms\ for the equivalent width
of $10^3$ \kms.
The \HH\ column density in the nuclear disks would be 
$\Sigma_{\rm H_{2}} = 1\times 10^{25}\, m_{\rm H_{2}}\, \persquarecm = 2\times 10^5$ \Msol\ \persquarepc\
if the line width in the disk is 300 \kms, 
which is half of the total line width due to disk rotation and is an upper limit to the
local line width in the disk.
The high column density is in order-of-magnitude agreement with what we inferred 
from the dust emission alone in \S\ref{s.result.continuum.tau_Td}, as it should be. 
The high column density $N_{\rm H_2} \sim 10^{25}$ \persquarecm\ 
suggests that the volume density in the nuclear disks is high,  
$n_{\rm H_2} \gtrsim 10^{4.5}$ \percubiccm\ 
for a volume filling factor of 1 and a line-of-sight depth of $\lesssim 100$ pc for the disks. 
The high density of the molecular gas is in agreement 
with the observations of molecular lines with high-critical densities 
\citep[e.g.,][]{Wiedner02,Imanishi07}
and is also in line with
models of \ion{H}{2} regions and supernova remnants \citep{Anan00,Parra07}.

A more elaborate model that also employs optically thick CO emission and explains
the lower CO equivalent width in the nuclear disks is that
the ISM has a temperature gradient in such a way that the gas and dust are hotter inside and cooler 
near the surface of the disk.
This reduces the CO emission through self-absorption and hence reduces the equivalent width
of the line, because the CO line has higher optical depth than the continuum, $\tau_{860} \sim 1$.
The absorption feature in the CO(2--1) line toward the west nucleus, observed by  \citest{Downes07}, 
suggests that this is a part of the reason for the low equivalent width.
It is quite possible that the disks have not only the temperature gradient but also 
a density gradient with the center having higher column and volume densities, 
as hinted by the compact dust emission.

\subsection{Energy Source}
\label{s.discuss.energy}
\subsubsection{Sources in the Nuclear Region}
Figure \ref{f.vlbisources} shows the centimeter radio sources and
the submillimeter continuum in the nuclear region.
Relative positions of various centimeter sources in the figure are based on observations
but the registration of centimeter and submillimeter data is made, 
as noted in \S \ref{s.result.continuum.positions},
with an assumption that the two nuclear sources seen in both wavelengths are the same. 
Specifically, the registration made the submillimeter peak coincide with the
median centroid of the supernova-related sources (hereafter SNe)
in the brighter west nucleus.
One can alternatively use the peak of diffuse centimeter emission for the registration,
because the offset between the two nuclei in that emission,
($\Delta\alpha, \Delta\delta$) = (0\farcs96, \minus0\farcs15)  \citep{Baan95},
also agrees with the offset measured in submillimeter.
However, the western peak of the centimeter diffuse continuum 
(green circle in Fig. \ref{f.vlbisources}) is only 40 mas
from the center of the SNe distribution (red circle in Fig. \ref{f.vlbisources}) 
according to the relative astrometry of
the diffuse emission with respect to the SNe by \citet{Rovilos03}.
Thus our centimeter--submillimeter registration in Fig. \ref{f.vlbisources} is
probably accurate to \about0\farcs04.

Neither of the two OH megamasers in the west nucleus 
\citep[shown in blue in Fig. \ref{f.vlbisources}]{Lonsdale98}
coincides with the submillimeter continuum peak,
according to the registration made above.
Of particular interest is the southern maser source called W2 in \citet{Lonsdale98}.
It has a large velocity gradient,
$dV/dr \approx 18.7$ \kms\ \perpc\ in a \about6 pc diameter region,
and may trace a heavily obscured AGN of \about$1.7 \times 10^{7}$ \Msol\ \citep{Rovilos05}. 
Whether it is an AGN or not, 
our submillimeter map and its registration in Fig. \ref{f.vlbisources} 
suggest that it is unlikely to be the main heating source of the
submillimeter-emitting dust.
We checked if the submillimeter data could be registered with the centimeter data
using the OH megamasers and found it unlikely.
Among the four OH megamasers in the nuclear region, 
the pair between the southern source in each nucleus has the separation that is
closest to the one between the submillimeter peaks.
This maser separation, ($\Delta\alpha, \Delta\delta$) = (0\farcs89, \minus0\farcs15), is in worse
agreement with the submillimeter separation (0\farcs96, \minus0\farcs18)
than the supernova-based separation (0\farcs94, \minus0\farcs18)
and the abovementioned separation in diffuse centimeter continuum.
Moreover, the shift needed for the registration is larger 
if the southwest megamaser should coincide with the west
submillimeter peak.
Furthermore, the elongation in the northeast--southwest direction seen both in the
submillimeter emission and the supernova distribution of the east nucleus
suggests that these two, rather than the submillimeter and OH emission, should be
correlated.
Thus, although the final conclusion should be made from more accurate astrometric observations,
we favor the registration in Fig. \ref{f.vlbisources} and the SNe--submillimeter connection 
over the megamaser--submillimeter connection.
The maser emission off the west nuclear disk may be collisionally excited by 
a bipolar wind or outflow from the disk.

\subsubsection{Supernovae and Dust Emission}
The spatial distribution of the submillimeter continuum is expected to follow the
surface density of supernovae under certain circumstances that include
the dominant energy source being a dust-enshrouded starburst (see Appendix \ref{a.Id-Sigma_sn}).
We show  the smoothed distribution of the SNe in Figure \ref{f.smoothedSNe}.
In Fig. \ref{f.smoothedSNe}a  the  smoothing kernel is a 0\farcs5 Gaussian 
that is about the same as the PSF of our low-resolution
data (Fig. \ref{f.low}) and the PSF of the mid-IR images of \citest{Soifer99}.
As noted earlier,
the eastern nucleus has an oval shape elongated in the same NE-SW direction
in the supernova distribution, 860 \micron\ continuum (see the deconvolved shape in Table \ref{t.datasets}),
and mid-IR continuum, among others. 
This agreement is suggestive of the dust being heated mainly by the starburst,
although there may still be room for an accreting black hole to hide within or behind the starbursting disk.
In contrast to the east nucleus, the brighter west nucleus shows different shapes in the 
submillimeter continuum and
in the supernova distribution (see Figs. \ref{f.smoothedSNe}b and  \ref{f.vlbisources}). 
The dust continuum is more compact and the starburst is much more elongated
in the east-west direction at the 0\farcs23 resolution.
This bright 860 \micron\ core has a minimum luminosity of $2\times10^{11}$ \Lsol, 
calculated in \S \ref{s.result.continuum.luminosity}.

The reason for the different spatial distribution between the submillimeter continuum
and SNe must be either
that many massive stars and their remnants are missed in the current VLBI observations at the
densest region or that there is a dust-heating source other than massive stars in the center of the
west nuclear disk.
The former could be because of  free-free absorption, radio quiet supernovae,
younger stellar population at the center of the W nucleus than the one around it and in the E
nucleus,
or
more/less dust and gas in the nucleus/disk of Arp 220 W than its star formation implies.
The non-stellar heating source, if it exists, must be an AGN.

Quantitatively, 
the west nucleus has three of the four new radio sources, i.e., radio supernovae, found
by \citet{Lonsdale06} in the merger over a period of one year. 
Two of them are in the center of the west nucleus,
and \citet{Parra07} pointed out that more core-collapse supernovae are probably left undetected
because the detected ones are most likely type IIn that are radio-luminous and rare.
It may therefore be possible that there is a stronger concentration of supernovae resulting from
vigorous star formation at the center of the west nuclear disk than is visible in the 
existing VLBI observations.
The starburst that caused the observed supernovae can provide the entire luminosity
of the merger even before the correction for non-type IIn
SNe \citep{Smith98,Lonsdale06,Parra07}, although such an estimate involves 
assumptions on the stellar mass function and starburst history, which will be difficult to verify.

\subsubsection{Constraints from Luminosity-to-Mass Ratio and Luminosity Surface Density}
\label{s.nulcei.energy_source.L/M_and_Sigma_L}
Among quantitative parameters that could constrain the energy source from our data
are luminosity-to-mass ratio and luminosity surface density.
A lower limit to the luminosity-to-mass ratio within $r=40$ pc of the west
nucleus is about $4\times10^{2}$ \Lsol/\Msol\ according to the parameters estimated in 
\S \ref{s.result.continuum.luminosity} and \S \ref{s.nuclei.dynamical}.
The ratio would be as high as $10^3$ \Lsol/\Msol\ 
if the nucleus has a luminosity of $10^{12}$ \Lsol\ as shown possible depending on
its size and optical depth.
The $L/M$ ratio, which depends on stellar mass function, may be as high as \about$1\times10^{3}$
\Lsol/\Msol\ for a very young starburst  \citep[age $< 10$ Myr, see, e.g.,][]{Leitherer99}.
Its more robust upper limit is \about$1\times10^{4}$ \Lsol/\Msol\ 
for a pure population of 100 \Msol\ stars.
The high ratios in the range of ($10^{3}$--$10^{4}$) \Lsol/\Msol\ are observed in 
super star clusters (SSCs) such as the one in NGC 5253 studied by \citet{Turner00}
and the Quintuplet and other clusters in \citet{Figer99}.
For comparison, an AGN fueled at the Eddington rate has a luminosity to black-hole
mass ratio of $4\times10^{4}$ \Lsol/\Msol.
If the $L/M$ ratio is close to our lower limit, then it is quite possible,  
as far as this parameter is concerned,
that the main luminosity source of the nucleus is an intense starburst 
producing many super clusters.
The $L/M$ ratio for the $10^{12}$ \Lsol\ case, on the other hand, 
is close to the largest a starburst can have, 
considering that the observed dynamical mass includes everything in the region, 
such as young and old stars, 
massive black holes (if any), and the gas and dust of high column densities.
We can therefore infer that the west nucleus, if it has a luminosity of \about$10^{12}$ \Lsol,
must have either an energetically significant AGN or
an extreme starburst that is compact ($r\leq40$ pc), young (age $\lesssim 10$ Myr),
and maybe biased toward high mass stars.

The luminosity surface density of the west nucleus is $10^{7.6}$ \Lsol\ \persquarepc\ or larger
in the central 80 pc.
The lower limit is already among the highest for infrared-luminous starburst galaxies 
\citep[][their Table 4]{Soifer01}.
If the west nucleus has a luminosity of $10^{12}$ \Lsol\ within 50 pc diameter, 
then the parameter reaches $10^{8.7}$ \Lsol\ \persquarepc.
Obviously, a luminous AGN would have no problem to achieve this level of luminosity surface density.
At the same time, there are known stellar systems that could also achieve the high values.
Some of the youngest super star clusters have core radii of 0.1--1 pc, luminosities of \about$10^{7}$--$10^{9}$ \Lsol, 
and ages less than 10 Myr \citep{Figer99, Turner00}.
Taking a luminous SSC of $10^{9}$ \Lsol\ as a yard stick, there would be
200 and 1000 of them, respectively, for the ($L$, $d$)=($2\times10^{11}$ \Lsol, 80 pc) and
($10^{12}$ \Lsol, 50 pc) cases mentioned above if the SSCs are the main luminosity source.
Their mean spacing would be 11 pc and 4 pc, respectively.
The clusters would probably overlap in their outskirts,
since half of the \about7000 O stars in the most luminous SSC in NGC 5235, 
used as the yard stick, are in the central 5 pc \citep{Turner04}.
Alternatively, many of the clusters may have already been disintegrated to form a single concentration 
of massive stars in the galaxy nucleus, considering the short dissolution time of SSCs \citep{Fall05}.
It could also be that  a smaller number of heavier cluster(s) was formed rather than hundreds of SSCs. 
This possibility cannot be excluded on the grounds of
the number density of massive stars required to produce the luminosity surface density of Arp 220 W,
since higher number densities are observed in SSCs.

\subsubsection{Star Cluster v.s. AGN}
As seen above, the parameters of the luminous west nucleus are in or close to the limit
explicable by a star cluster or clusters.
The starburst model may explain some of the observations that appear to favor an AGN but 
it certainly has some difficulties.
The extreme luminosity surface density even for a ULIRG
is expected to result in photodestruction of PAH particles as was observed around
the model SSC in NGC 5253 \citep{Beirao06}. 
This may be part of the reason for the merger's low PAH emission to 850 \micron\ continuum ratio
that was taken as evidence for an AGN \citep{Haas01}.
The excess dust emission at the center of the west nucleus compared to the number of
supernovae and young supernova remnants is reminiscent of the abovementioned SSC in NGC 5253
whose predominantly thermal radio spectrum is attributed to a very young starburst with few supernovae
\citep{Beck96}.
On the other hand, as \citet{Haas01} emphasized, a cluster of stars (or SSCs) with an extent of a few 10 pc or larger
would be more difficult to enshroud with dust than an accreting black hole.
The cluster model may therefore be incompatible with the deeply enshrouded but energetically
dominant component used by \citet{Spoon04} and \citet{Dudley97} to model the mid-IR spectrum.
Another difficulty in the cluster model is that it needs to form hundreds of SSCs
or equivalent  high-mass stars at high volume density in the short time of $< 10$ Myr.
The short time scale is necessary because the luminosity-to-mass ratio of a cluster declines once
massive stars start to die and also because gas and
dust in the nucleus would be dispersed by the energy injection from the starburst.
Although SSCs tend to form in groups \citep{Zhang01}, we do not yet know the mechanism for such 
a compact and intense burst of SSC formation or even whether it is possible.

A massive black hole of \about$10^{8}$ \Msol\ can generate (0.2--1)$\times10^{12}$ \Lsol\
via mass accretion at a sub-Eddington rate, as it is believed to do in quasars.
The above-mentioned constraints on mass-to-luminosity ratio and 
luminosity surface density of the west nucleus can be easily met with a buried quasar.
The weaker concentration of supernovae and supernova remnants toward the center
of W than the concentration of the submillimeter continuum and bolometric luminosity
can be explained if a large part of the luminosity is due to an AGN.
There is enough material, $N_{\rm H_2} \sim 10^{25}$ \persquarecm, to block the X-rays,
and it is easier to enshroud an AGN than a million O stars.
An AGN with a high-covering factor could explain the large equivalent width of 
the Fe K$\alpha$ line of Arp 220 \citep{Iwasawa05}.
Regarding the formation mechanism, it is not surprising that the remnant nucleus of 
one or both of the merged galaxies has a massive black hole, 
since most if not all galaxies are thought to harbor a massive black hole at the center.
The massive black hole in Arp 220 W would be at the center of the 
nearly edge-on nuclear disk of gas and dust,
having abundant gas in its vicinity to accrete,
and might have been significantly growing through the increased supply of fuel.
The number of supernovae and the amount of free-free emission that appear 
to suggest a starburst large enough to provide the entire luminosity
of Arp 220 is a problem for the luminous-AGN model.
Our observations suggest that neither of the two OH megamasers about 50 pc off the 
luminosity center marks the main luminosity source even if they are associated with AGNs.

Taking everything above into account, 
a luminous AGN at the center of the west nucleus, proposed most recently by \citest{Downes07},
seems to have fewer difficulties than 
the SSC-cluster model if the luminosity of the bright 860 \micron\ core is significantly larger
than the lower limit we set.
If, on the other hand, the luminosity is close to the observational lower limit, then the
majority of the luminosity of the west nucleus is emitted from the nuclear disk rather
than the core.
In this case the heating source for the disk, as well as for the core, can be a starburst.
In other words, the spatial distribution of luminosity within the west nucleus is found more
centrally peaked than known before (including \citest{Downes07}), 
but the parameters of the concentration obtained so far
are not enough to conclude the existence of an energetically-dominant AGN.
Really robust evidence is needed here considering the number of previous studies that 
have concluded one way or the other about this issue.

\subsection{Prospects} 
Further observations will let us set tighter constraints on the energy sources 
and their spatial distributions.
For example, higher resolution observations at around 860 \micron\ will better 
determine the luminosity-to-mass ratio.
Such observations may soon become possible through the ongoing eSMA project
that combines the SMA with the neighboring JCMT and CSO telescope.
Observations at even shorter submillimeter wavelengths, where dust opacity is higher, 
will better constrain the dust temperatures in the nuclear and outer disks and
will thereby determine their relative contributions to the total luminosity.
They also let us search for a compact and bright core in the east nucleus.
The luminosity estimate from  dust thermal emission in the submillimeter wavelengths complements
other methods, because it bypasses the number of ionizing photons or supernovae to estimate luminosity  
and  also because it is less hindered by foreground extinction than infrared-based methods.
In addition, various molecular lines in millimeter and submillimeter will tell us the physical 
properties of the ISM and alert us to any anomaly in the nuclei.
Ongoing projects in these directions, further modeling of the structure and radiation of the ISM,
and the eventual arrival of ALMA will tell us the heating sources not only in Arp 220 but also 
in ULIRGs in general.
As an example, high-resolution photometry at a \about10 pc resolution across the ALMA 
observing bands will allow us to locate a Compton-thick AGN of $10^{12}$ \Lsol\ as a source
of a few 100 K, even if it is hidden in other wavelengths behind layers of absorbers.
The dynamical mass estimated from molecular lines
may reveal that its luminosity to mass ratio is too large for a starburst.
Finding many such sources may prove the long-hypothesized formation of 
quasars in luminous mergers \citep[and references therein]{Sanders88,Hopkins06}.

\section{Summary}
\label{s.summary}
The ultraluminous merging galaxy Arp 220 has been imaged using
the Submillimeter Array (SMA)
at 0\farcs2--0\farcs3 (\about100 pc) resolutions in the
CO(3--2) line and 860 \micron\ continuum.
We analyzed structure and properties of the submillimeter emission with
particular attention to the spatial distribution of bolometric luminosity.
The luminosity distribution allows us to constrain the energy source of Arp 220.
Our main results are as follows.

1. 
Both the CO(3--2) line and the submillimeter continuum peak at or around the
two nuclei of the merger.
The continuum emission is mostly from compact (a few 0\farcs1) regions
at the two nuclei.
The CO gas is more extended than the continuum at the nuclei, and it also
has a component with a total extent of \about3\asec (1 kpc)
that encompasses the double nucleus.
The overall distribution of CO and dust emission is in agreement with the
observations at lower frequencies.

2. 
The peak brightness temperature exceeds 50 K in both the CO(3--2) line
and the submillimeter continuum emission, revealing the presence of warm gas
and dust in the center of the merger.
The west nucleus has the strongest continuum emission that is consistent with 
a Gaussian of FWHM 0\farcs15 (50 pc) and peak brightness temperature $1.6\times10^{2}$ K
or a uniform-brightness disk of diameter 0\farcs23 (80 pc) 
and a brightness temperature of  $0.9\times10^{2}$ K.

3.
The dust opacity at both nuclei is on the order of unity at 860 \micron, 
according to the analysis of flux ratios to 1.3 mm and to 25 \micron.
Both the 860 \micron\ opacity and low equivalent widths of the CO(3--2) line
at the two nuclei (\about$10^{3}$ \kms) 
suggest a high column density of gas and dust toward the nuclei, of the order of
$N_{\rm H_{2}} \sim 10^{25}$ \persquarecm.
The equivalent-width analysis and the CO(3--2) to CO(2--1) intensity ratio close to 1
suggest that the CO(3--2) line is optically thick and the molecule is thermalized 
at least up to the J=3 level in the nuclear disks.
The gas there must be also dense, $n_{\rm H_2} \gtrsim 10^{4.5}$ \percubiccm.

4.
The bolometric luminosity of the bright 50--80 pc core within the west nucleus is at least $2\times10^{11}$ \Lsol\ 
and may be $10^{12}$ \Lsol\ or higher depending on the opacity and structure of the source. 
Our blackbody lower limit is an order of magnitude larger than that from existing 1.3 mm observations.
The east nucleus has a lower limit to the bolometric luminosity of $5\times10^{9}$ \Lsol.
It also may be more luminous if its 860 \micron\ emission is optically thin.

5. 
The sharp velocity shift across each nucleus is confirmed in the CO(3--2) velocity field,
and so is the overall rotation of the larger (\about kpc) gas disk.
Each nucleus has a line-width of \about500 \kms\ and 
a dynamical mass of \about$1\times10^{9}$ \Msol\ within a radius 
of \about100 pc if the gas motion is due to rotation, as seems most likely.
The mean mass densities in the nuclei are on the order of $10^2$--$10^3$ \Msol\ \percubicpc,
and are comparable to those of galaxy centers at the same scale.
The overall merger system has a full CO line width of \about1100 \kms\ 
and does not show extremely high velocity CO at our sensitivity.
The overall structure and kinematics are consistent with what has been proposed, 
a binary system of counter-rotating nuclear disks encompassed in a larger disk.

6. 
Morphological agreement between dust continuum and surface density of supernova-related
VLBI sources in the east nucleus is suggestive that the dust is mainly heated by a starburst.
In contrast, the west nucleus has more compact dust emission than the supernova distribution.
The compact hot core has a luminosity-to-mass ratio of $\gtrsim 4\times10^{2}$ \Lsol/\Msol\
and a luminosity surface density of $\geq 10^{7.6}$ \Lsol\ \persquarepc\ in the central 80 pc,
with a possibility of actually having \about5 times larger values than these lower limits.
These parameters, at least the lower limits, are similar to
those in young super star clusters and are close to the highest for a starburst.
Hence we can not yet rule out an extreme starburst equivalent to hundreds of young SSCs.
A buried luminous AGN accompanied by a starburst
can also account for the luminosity and other observed parameters of the west nucleus,
and it becomes more plausible in higher luminosity.
Further observations as well as modeling are needed to determine the luminosity
distribution and decide the source. 
High resolution imaging of submillimeter emission is a promising way to accomplish that.
The OH megamaser in the west nucleus is off the peak of dust continuum and luminosity,
and is hence unlikely to mark the position of a dominant heating source such as an AGN.

\begin{figure}[t]
\epsscale{0.5}
\plotone{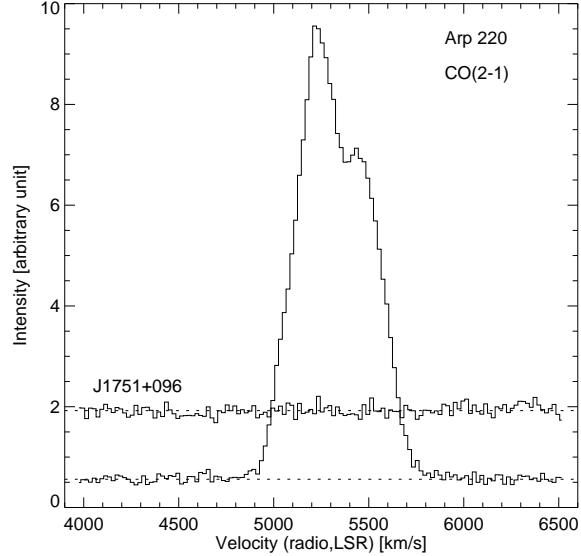}	
\epsscale{1.0}
\caption{ \label{f.co2-1spectrum}
CO(2--1) spectrum of Arp 220 obtained from the SMA observations of
\citet{Matsushita07}.
The data cube from which the spectrum is obtained has a velocity resolution of 20 \kms\
and is convolved to a resolution of 5\arcsec\ (FWHM). 
Continuum is not subtracted.
The horizontal dotted line shows the mean intensity of the outermost 40 channels of the spectrum
(i.e., 20 from each edge).
We adopt, on the basis of this spectrum, the channels outside of the velocity range of 4800--5900 \kms\ as continuum channels.
The CO spectrum has a peak signal-to-noise ratio of 147 and sets an upper limit to 
high-velocity CO emission ---
such emission in the adopted continuum channels has no more than a few percent of the peak 
line intensity of the galaxy at this resolution.
Also plotted is the (flat) spectrum of the quasar J1751+096 observed during the same track
and was processed in the same way as Arp 220. 
It shows the flatness of the line-free passband.}
\end{figure}

\begin{figure}[t]
\epsscale{0.5}
\plotone{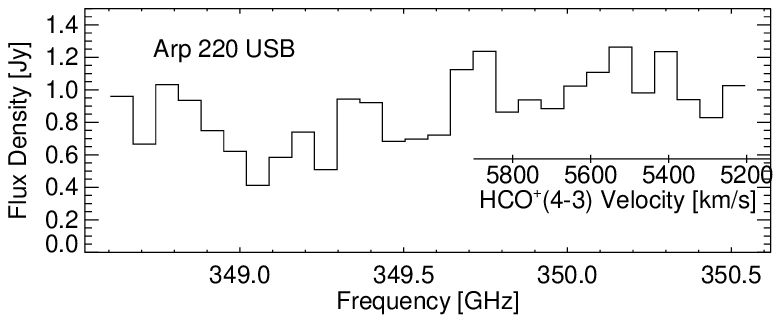} \\	
\epsscale{0.5}
\plotone{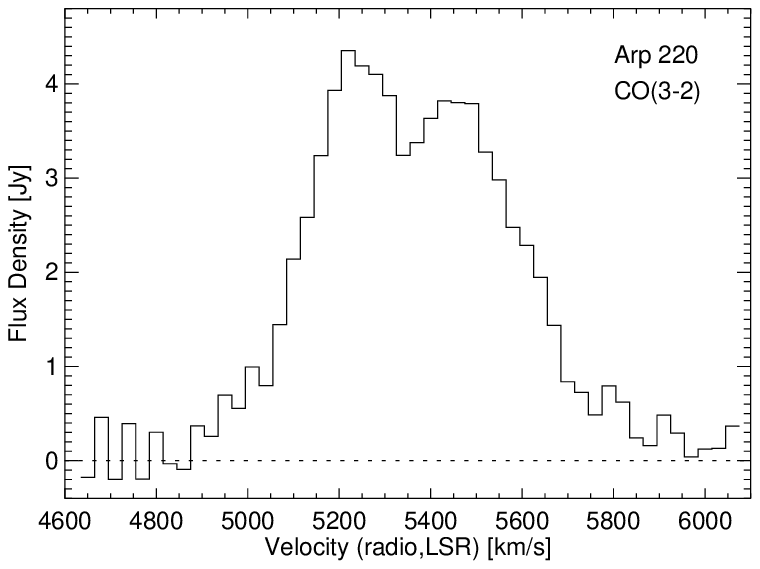}  
\epsscale{1.0}
\caption{ \label{f.spectra}
(top) 
A spectrum of Arp 220 in the central 3\arcsec\ of the merger,
taken from our USB data at 0\farcs5 resolution. No continuum subtraction is made.
The inset velocity axis is for the HCO$^{+}$(4--3) line.
(bottom)
CO(3--2) spectrum of Arp 220 in the central 3\arcsec\ of the merger,
taken from our LSB data cube of 0\farcs5 resolution.
Continuum has been subtracted.
The line has a total flux of  $2.2\times10^3$ Jy \kms\
in the velocity range of 4800--5900 \kms.}
\end{figure}

\begin{figure}[t]
\epsscale{0.315}
\plotone{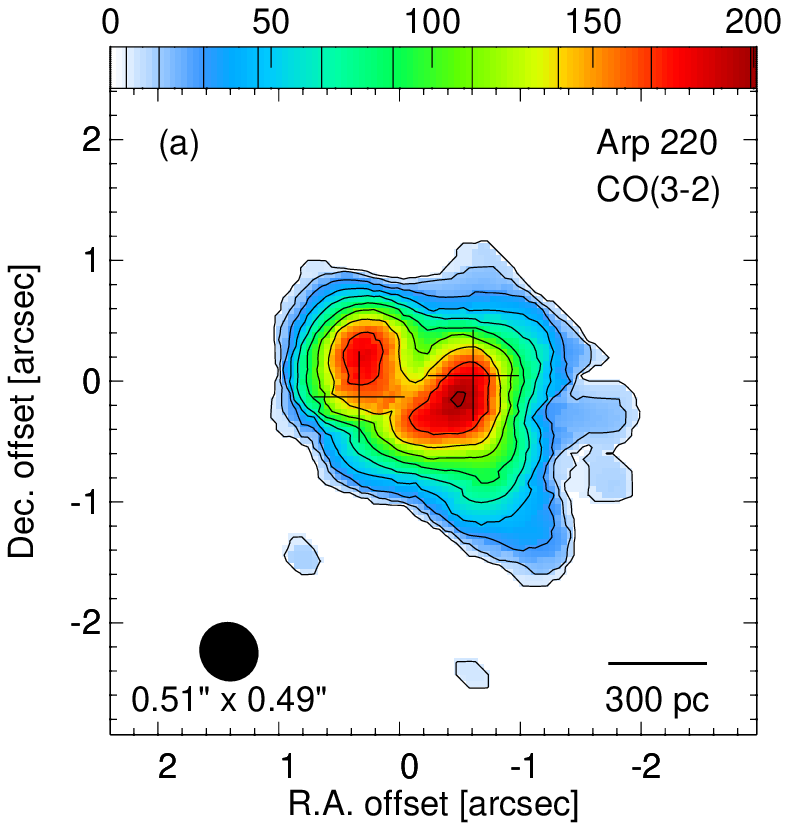}	
\epsscale{0.315}
\plotone{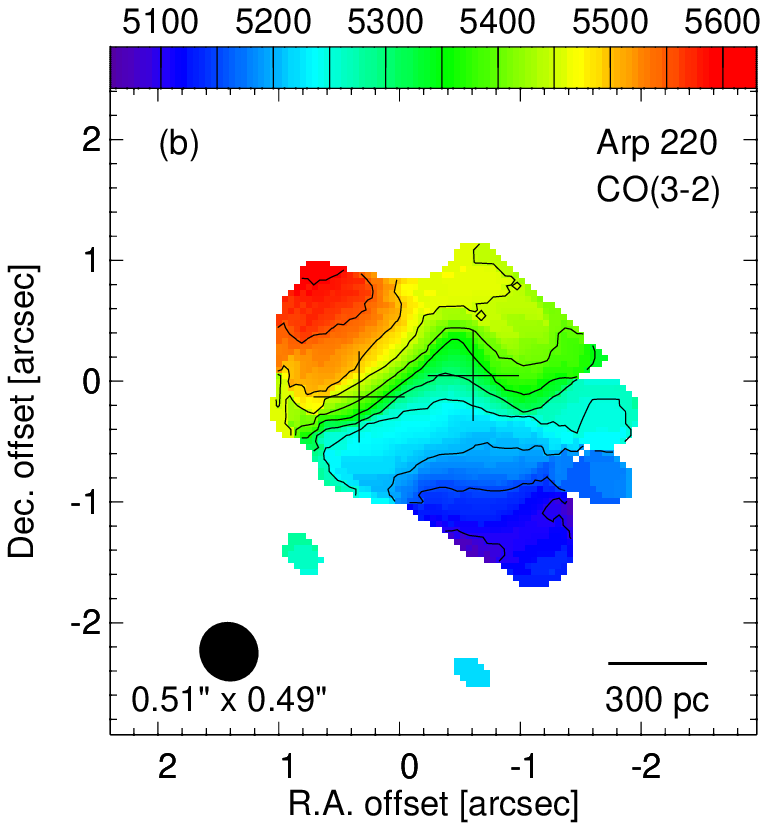}	
\epsscale{0.315}
\plotone{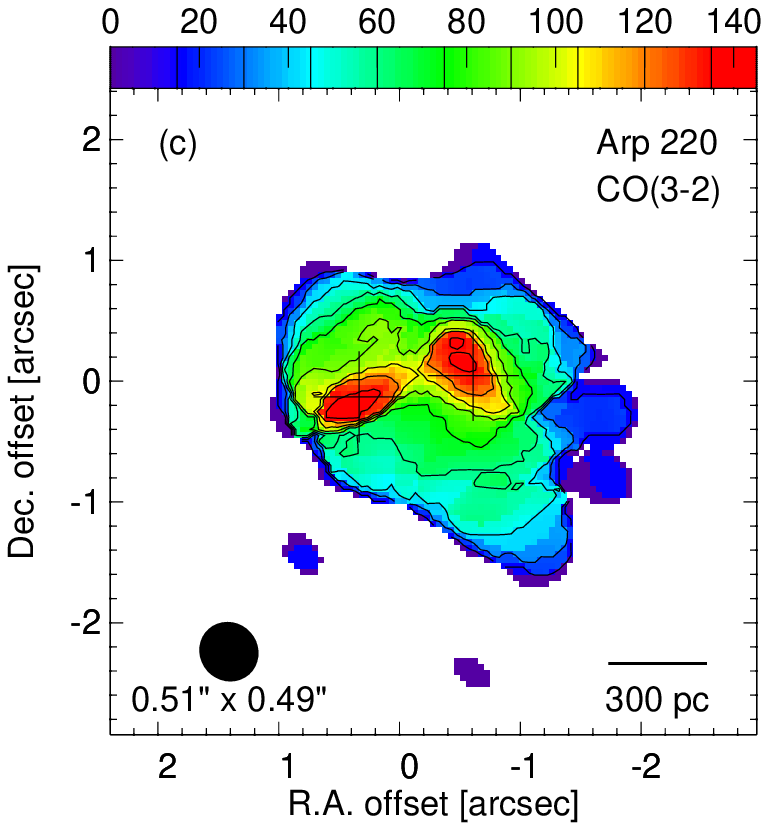}	
\epsscale{0.315}
\plotone{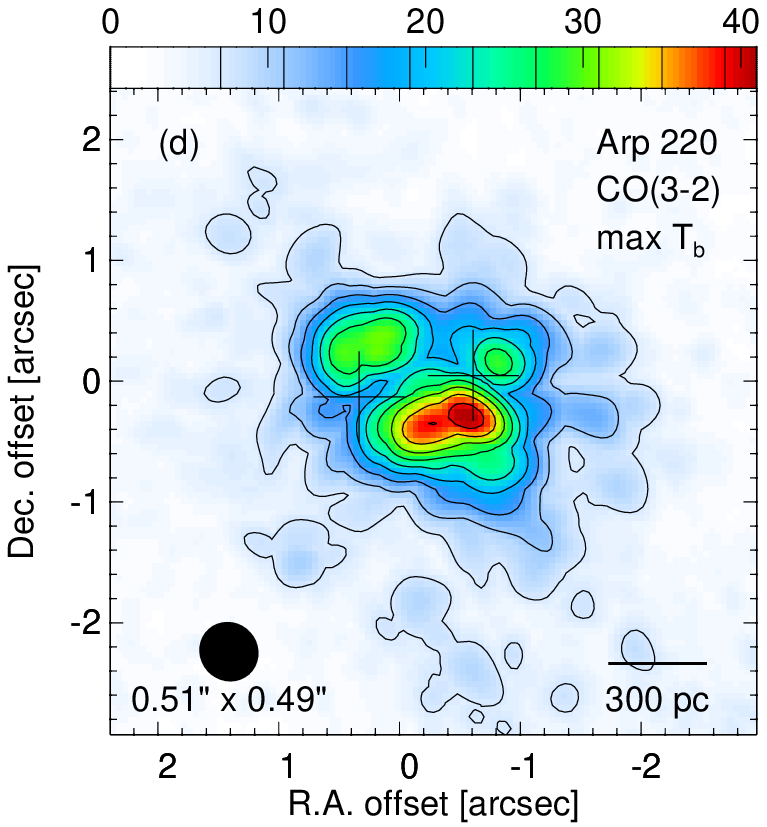}	
\epsscale{0.315}
\plotone{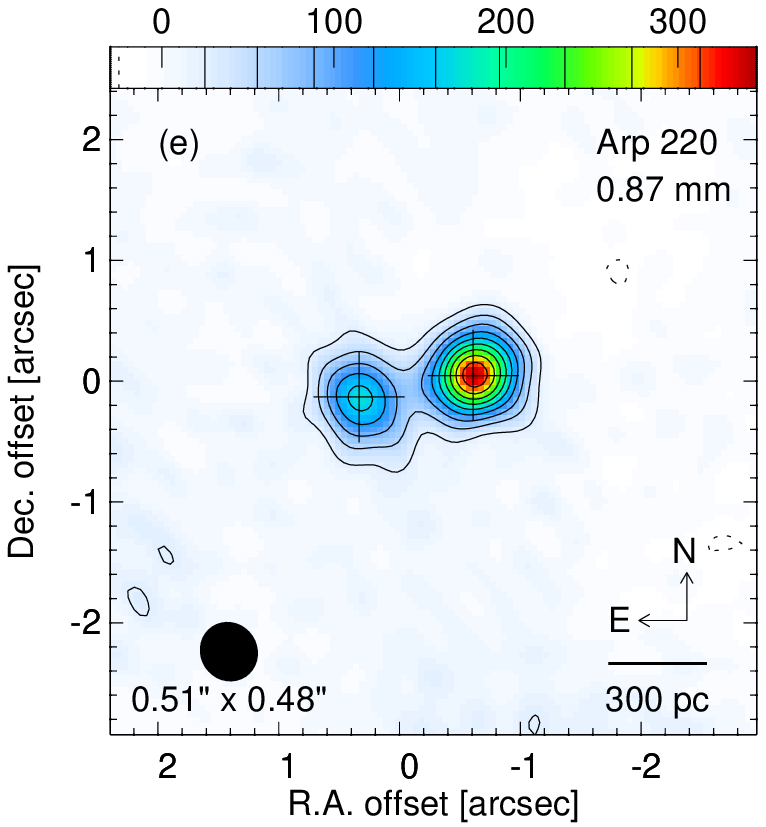}	
\epsscale{1.0}
\caption{ \label{f.low}
SMA 0\farcs5 resolution images of the CO(3-2) line and 345 GHz continuum in Arp 220.
(a) CO(3--2) integrated intensity (i.e., the 0th moment). The $n$th contour is at $5\times n^{1.6}$ Jy \perbeam\ \kms.
(b) CO(3--2) mean velocity (i.e., the 1st moment). Contours are in steps of 50 \kms.
(c) CO(3--2) velocity dispersion (i.e., the 2nd moment). Contours are in steps of 15 \kms.
(d) CO(3-2) peak brightness temperature calculated using the Rayleigh-Jeans approximation.  
Contours are in steps of 4 K starting from 7 K.
(e) 345 GHz continuum. 
The $n$th contour is at $25\times n^{1.15}$ mJy \perbeam. 
The first contour is at 2.5 $\sigma$ and negative contours are dashed.
The two plus signs are at the continuum peaks whose positions are in Table \ref{t.positions}. 
The positional offsets are measured from the phase tracking center listed in Table \ref{t.obsparam}. 
At the bottom-left corner of each panel is the FWHM size of the synthesized beam.
}
\end{figure}

\begin{figure}[t]
\plottwo{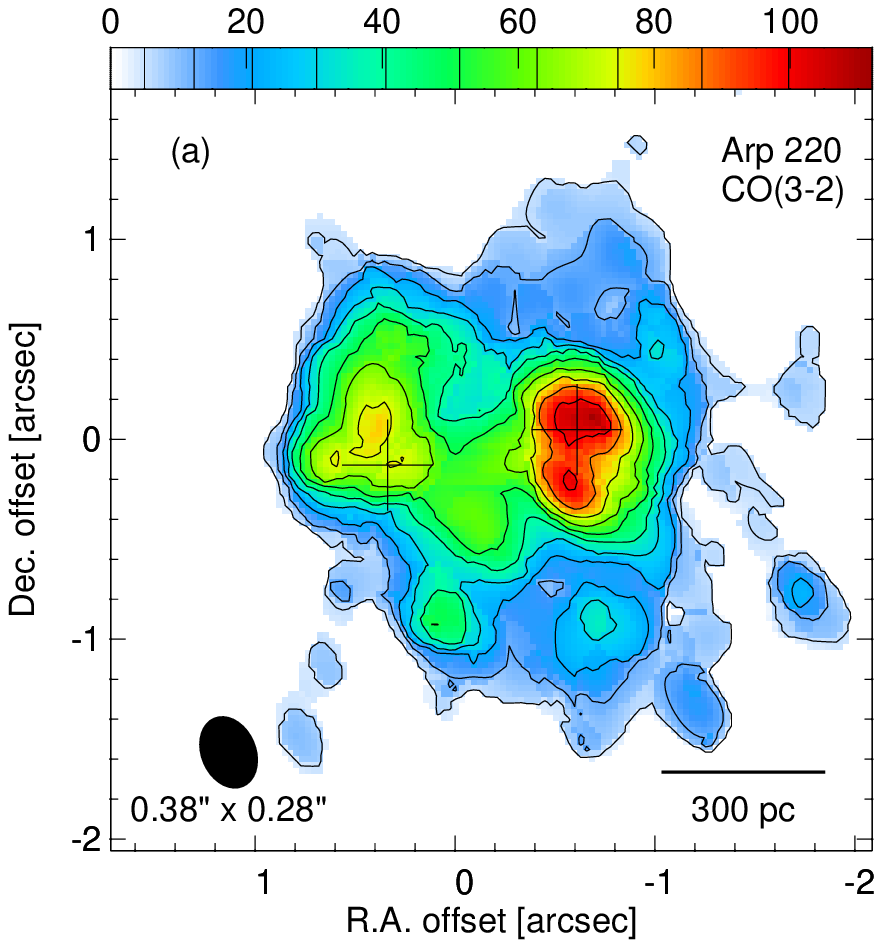}{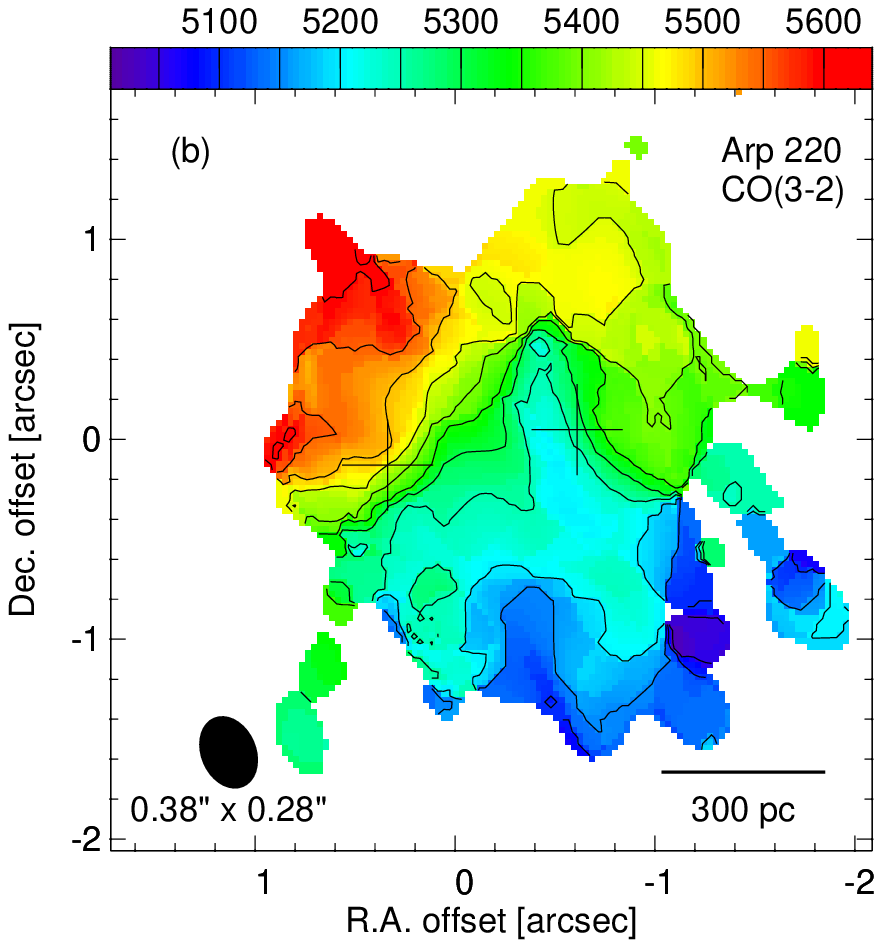} \\ 
\plottwo{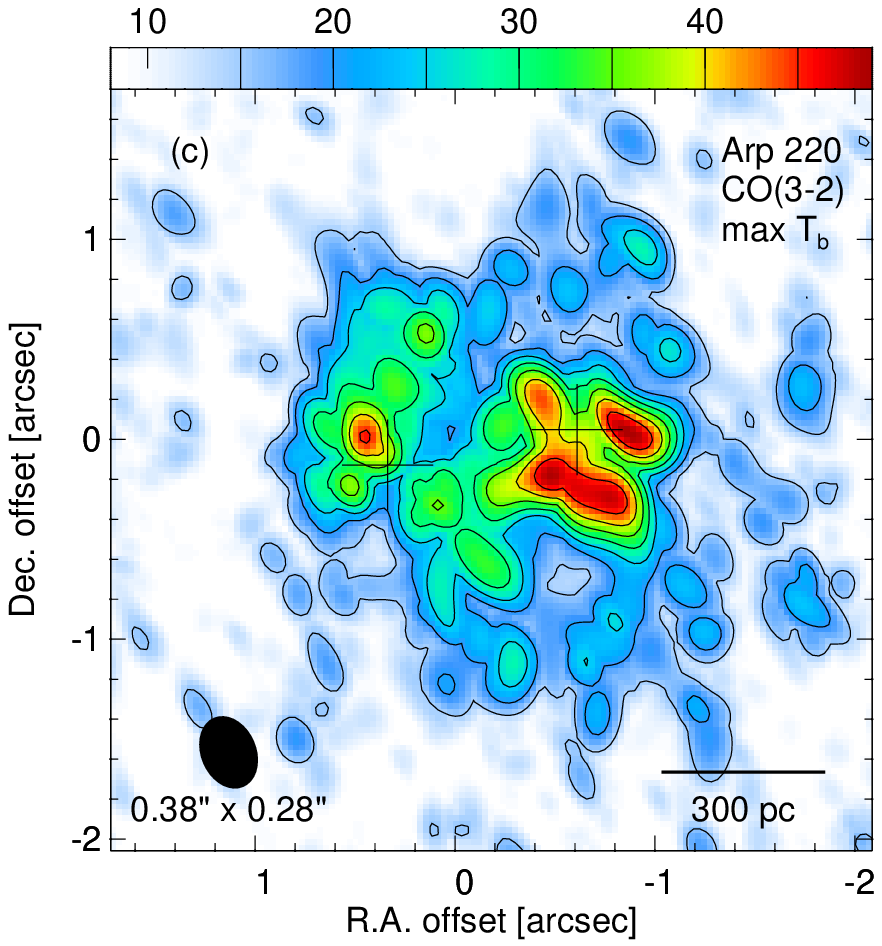}{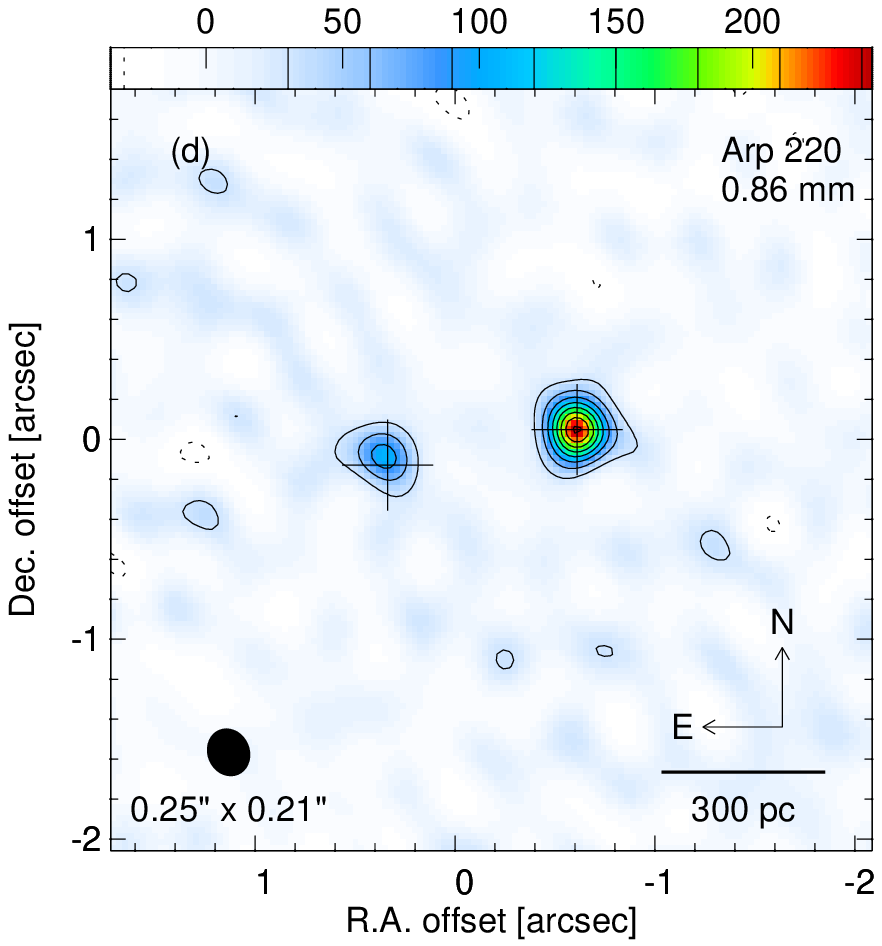}	
\epsscale{1.0}
\caption{ \label{f.high}
High resolution SMA images of the CO(3-2) line and 350 GHz continuum in Arp 220.
(a) CO(3--2) integrated intensity (i.e., the 0th moment). The $n$th contour is at $5\times n^{1.3}$ Jy \perbeam\ \kms.
(b) CO(3--2) mean velocity (i.e., the 1st moment). Contours are in steps of 50 \kms.
(c) CO(3-2) peak brightness temperature calculated with the Rayleigh-Jeans approximation.
 Contours are in steps of 5 K starting from 15 K.
(d) 350 GHz continuum. Contours are in steps of 30 mJy \perbeam\ (=3$\sigma$). 
Negative contours are dashed.
The two plus signs are at the continuum peaks whose positions are in Table \ref{t.positions}. 
The positional offsets are measured from the phase tracking center in Table \ref{t.obsparam}. 
At the bottom-left corner of each panel is the FHWM size of the synthesized beam.
}
\end{figure}

\begin{figure}[t]
\epsscale{1.0}
\plottwo{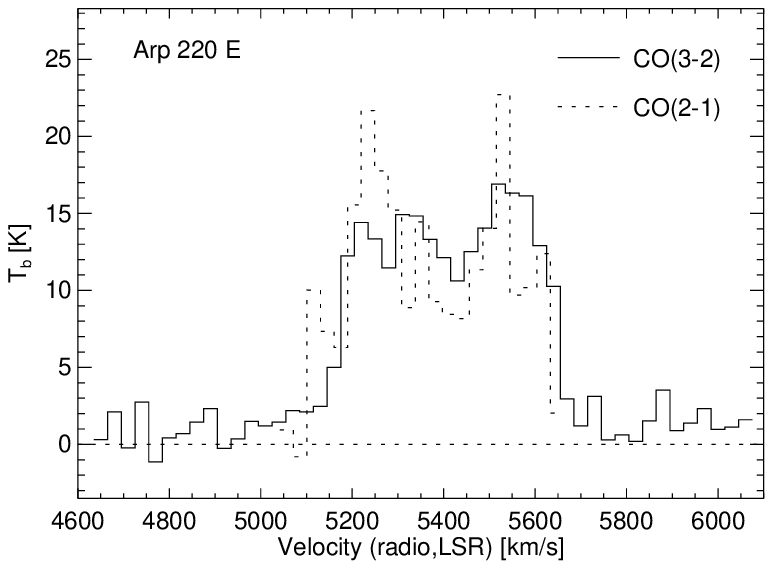}{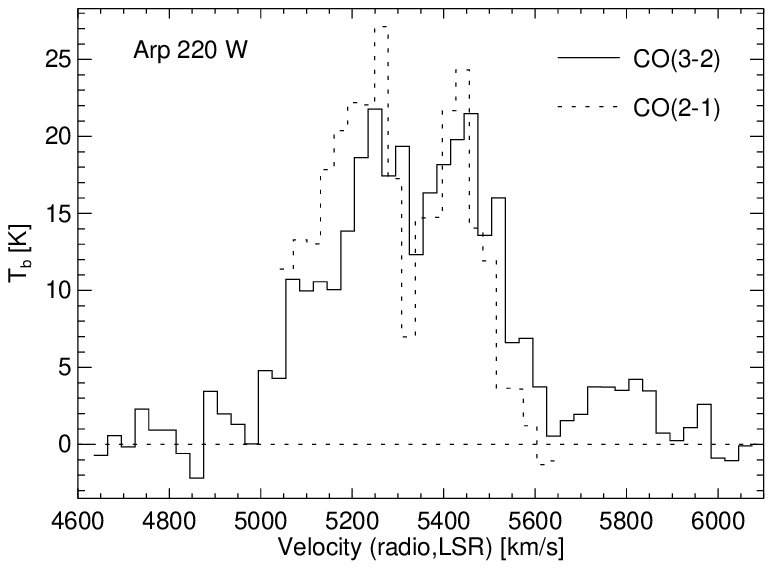}  
\caption{ \label{f.nucleus.spectra}
CO spectra of Arp 220 E (left) and Arp 220 W (right).
The CO(3--2) spectra are from our 0\farcs5 resolution data 
and the CO(2--1) from the 0\farcs57$\times$0\farcs52 data of \citest{Sakamoto99}.
Continuum has been subtracted. Brightness temperatures are from the Rayleigh-Jeans approximation.
}
\end{figure}

\begin{figure}[t]
\epsscale{0.8}
\plottwo{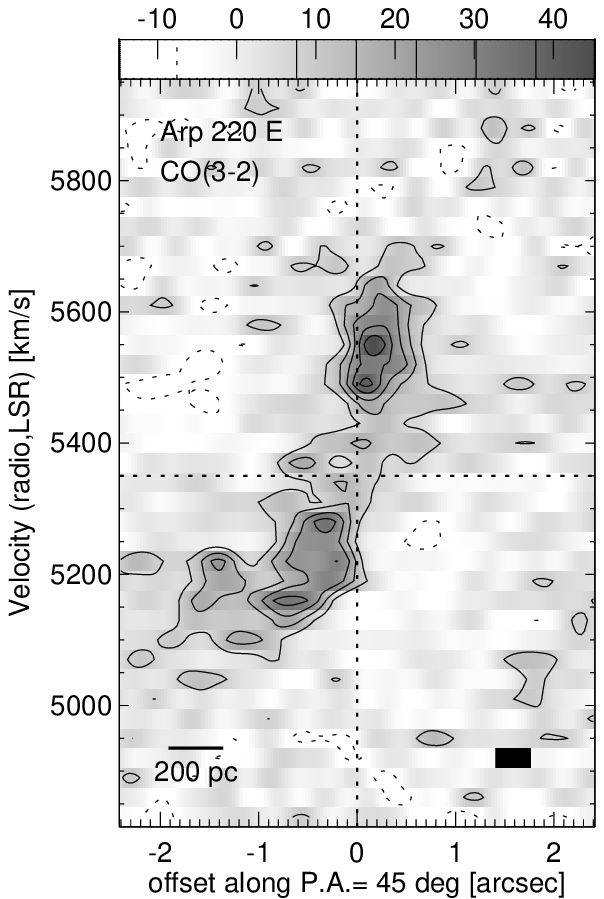}{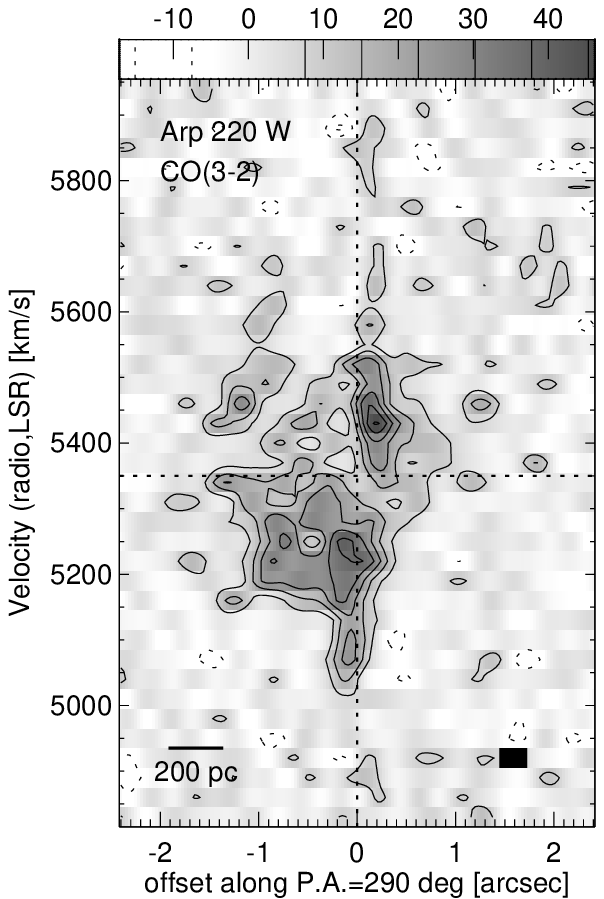}  
\caption{ \label{f.pv.nucleus}
Positions-velocity diagrams of CO(3--2) emission across Arp 220 E (left) and Arp 220 W (right).
Our 0\farcs33 resolution data with continuum subtraction are used.
Intensities are in the Rayleigh-Jeans brightness temperature \KRJ.
Contours are in 1.5 $\sigma$ steps with zero contours omitted and negative contours dashed.
The vertical dotted line is at each nucleus (see Table \ref{t.positions} for the coordinates). 
The horizontal dotted line at 4350 \kms\ is the approximate systemic velocity of Arp 220.
The black rectangle at the bottom right corner shows the spatial and velocity resolutions.
The spatial resolution along the cut is 0\farcs36 for E and 0\farcs28 for W.
}
\end{figure}

\acknowledgements
We are grateful to people in and around the SMA project for making our challenging observations possible.
We also thank Satoki Matsushita for the CO(2--1) data in Fig. \ref{f.co2-1spectrum}
and the referee for comments that improved the paper.
This research made use of 
the NASA/IPAC Extragalactic Database (NED),
the NASA's Astrophysics Data System (ADS), 
and the computer system in
the Astronomical Data Analysis Center (ADAC) of the National Astronomical Observatory of Japan.

{\it Facilities:} \facility{SMA}



\begin{deluxetable}{lcc}
\tablewidth{0pt}
\tablecaption{Arp 220 parameters  \label{t.galparm} }
\tablehead{ 
	\colhead{parameter}       &
	\colhead{value} &
	\colhead{ref.}
}
\startdata
luminosity distance  [Mpc] & 77.4 &  1\\
angular-size distance  [Mpc] & 74.5 &  1 \\
scale & 1$''$=361 pc & 1\\
$V_{\rm sys}$(radio, LSR) [\kms] & \about5350  & 2\\
$\log (L_{\rm 8-1000\; \mu m}/\Lsol)$  & 12.2 & 3
\enddata
\tablerefs{
1. NED;
2. \citet{Wiedner02};
3. \citet{Sanders03} 
}
\end{deluxetable}

\begin{deluxetable}{lc}
\tablewidth{0pt}
\tablecaption{SMA observation parameters  \label{t.obsparam} }
\tablehead{ 
	\colhead{parameter}       &
	\colhead{value} 
}
\startdata
position (J2000.0)      		& $\alpha$=\phs15\hr34\mn57\fs26 \\
				    		& $\delta$=+23\arcdeg30\arcmin11\farcs4\phn  \\	
field of view (LSB/USB)\tablenotemark{a}  &  35\asec\ / 34\asec \\						
center frequency  (LSB/USB)    & 339.58 / 349.58 GHz \\							
bandwidth\tablenotemark{b}               & 2.0 GHz (1732 \kms) \\
spectral resolution\tablenotemark{c}         	       		& 3.25 MHz (2.8 \kms) \\
gain calibrators                             & J1613+342, J1635+381 \\
primary flux calibrator                                 & Uranus \\
projected baselines for Arp 220      & 10.2 m -- 508.1 m
\enddata
\tablenotetext{a}{Full width at half maximum of the primary beam in each sideband.}
\tablenotetext{b}{Bandwidth in each sideband. The velocity coverage for the CO(3--2) line is in the parenthesis.}
\tablenotetext{c}{The velocity resolution for the CO(3--2) line is in the parenthesis.}
\end{deluxetable}


\begin{deluxetable}{cclccccccc}
\tabletypesize{\scriptsize}
\tablewidth{0pt}
\tablecaption{Log of SMA observations  \label{t.obslog} }
\tablehead{ 
	\colhead{No.}       &
	\colhead{UT date}  &
	\colhead{configuration} &
	\colhead{\# Ant.} &
	\colhead{$\tau_{225}$} &
	\colhead{$\langle T_{\rm sys, DSB}\rangle$} &
	\colhead{$L_{\rm baseline}$} &
	\colhead{$T_{\rm obs}$} &
	\colhead{$S_{\rm 345\; GHz}^{\rm J1613+342}$} &
	\colhead{$S_{\rm 345\; GHz}^{\rm J1635+381}$} 
	\\
	\colhead{ }       &
	\colhead{ }  &
	\colhead{ } &
	\colhead{ } &
	\colhead{} &
	\colhead{[K]} &
	\colhead{[m]} &
	\colhead{[hr]} &
	\colhead{[Jy]} &	
	\colhead{[Jy]}
	\\
	\colhead{(1)}  &
	\colhead{(2)}  &
	\colhead{(3)} &
	\colhead{(4)} &
	\colhead{(5)} &
	\colhead{(6)} &
	\colhead{(7)} &
	\colhead{(8)} &
	\colhead{(9)} &	
	\colhead{(10)}
}
\startdata
3 & 2004-06-11 &   1,8,9,11,14,17,18   &  7 &  0.07 & 335 & 25.8--352.9  & 4.6 & \nd & (2.41)  \\
5 & 2004-07-08 &   1,5,7,8,9,11,16,23 &   8  &  0.08 & 348 & 13.5--122.9      & 3.5 & 1.05 & 2.10 \\
6 & 2005-05-22 &  1,11,17,18,19,21    &   6  &  0.06 & 274 & 36.5--508.1      & 3.4 & 1.15 & 0.99 \\
7 & 2005-08-12 &   1,4,5,7,8,11,23             &   7    &  0.08 & 324 & 10.2--\phd68.7  & 2.4 & 1.33 & 0.71 
\enddata
\tablecomments{ 
(1) Track number.
(2) UT data in the $yyyy$-$mm$-$dd$ format.
(3) Array configuration. Antenna locations are listed \citep[see][for a map with the numeric keys]{Ho04}.
(4) Number of participating antennas.
(5) Zenith opacity at 225 GHz measured at the Caltech Submillimeter Observatory
next to the SMA.
(6) Median system temperature toward Arp 220.
(7) Range of projected baseline length for the galaxy.
(8) Total integration time on the galaxy (not necessarily with all the participating antennas, 
because of data flagging).
(9) and (10) Flux densities of the quasars used for gain calibration.
}
\end{deluxetable}

\begin{deluxetable}{lcccccl}
\tabletypesize{\tiny}
\tablewidth{0pt}
\tablecaption{CO and Continuum Parameters  \label{t.datasets} }
\tablehead{ 
         \colhead{parameter} &
         \multicolumn{2}{c}{CO(3--2) line} &
         \colhead{ } &
         \multicolumn{2}{c}{continuum} &
         \colhead{unit}
\\
	\cline{2-3} \cline{5-6} 
\\	
	\colhead{ } &         
	\colhead{low res.}       &
	\colhead{high res.}  &
	\colhead{ } &
	\colhead{low res.}  &
	\colhead{high res.} &
	\colhead{}	
}
\startdata
resolution (FWHM) & 0.51$\times$0.49 & 0.38$\times$0.28 & & 0.51$\times$0.48 & 0.25$\times$0.21 & arcsec$^2$ \\
linear resolution      & 180 & 120 &  & 180 & 80 & pc \\
center frequency & 339.6 & 339.6 &  & 344.6 & 349.6 & GHz \\
velocity resolution & 30 & 30 & & \nd & \nd & \kms  \\
bandwidth & \nd & \nd & & 1.5\tnm{a} & 2 & GHz \\
scaling factor to \Tb & 42.4 & 99.0 & & 42.1 & 189 & \KRJ/(Jy \perbeam) \\
r.m.s. noise & 38 & 51 & & 9.8 & 11.6 & mJy \perbeam \\
                      & 1.6 & 5.0 & & 0.41 & 2.2 & \KRJ\ \\
line flux in the central 3\arcsec & $2.2\times10^{3}$ & $1.6\times10^{3}$ & & \nd & \nd & Jy \kms \\
continuum in the central 3\arcsec & \nd & \nd & & 0.78 & 0.72 & Jy \\
peak integrated intensity & 1.8, 2.0 & 0.8, 1.1 & & \nd & \nd & $10^2$ Jy \perbeam\ \kms \\
                            & 7.8, 8.5 & 7.8, 11 & & \nd & \nd & $10^3$ K \kms \\  
peak intensity   & 0.98 & 0.46, 0.50 & & 0.18, 0.35 & 0.11, 0.24 & Jy \perbeam \\
                            & 42            & 46, 49 & & \phd7, 15 & 20, 46 & \KRJ\ \\
                            & 49            & 53, 57 & & 14, 22 & 27, 54 & K (Planck) \\      
CO(3--2) to (2--1) ratio\tnm{b} & 1.1, 1.0 & \nd & & \nd & \nd & \\                                                  
E, W line flux\tnm{c}               & 3.0, 3.2 & 2.8, 3.7 & & \nd & \nd & $10^2$ Jy \kms \\
E, W continuum flux density\tnm{b}   & \nd & \nd & & 0.20, 0.38 & 0.19, 0.36 & Jy \\
deconvolved size\tnm{d} (E)   & \nd & \nd & & 0.31$\times$0.25,\phd41 & 0.27$\times$0.14,\phd67 & arcsec$^2$, deg \\
deconvolved size\tnm{d} (W)   & \nd & \nd & & 0.28$\times$0.22,144 & 0.16$\times$0.13,109 & arcsec$^2$, deg \\
deconvolved peak intensity    & \nd & \nd & & 43, 91 & 52, 162 & K (Planck) \\
spectral index $\alpha$(230,345)\tnm{e}        & \nd & \nd & & 2.6$\pm$0.6, 2.7$\pm$0.5 & \nd & \\
line width at 20\% of maximum & 540, 600 & \nd & & \nd & \nd & \kms  \\
kinematical major axis  p.a. & \nd & \about45, \about290 & & \nd & \nd & deg \\
$V/R$ & \nd & 2, 6 & & \nd & \nd & \kms\ \perpc \\
$R$ & \nd & 11, 4  & & \nd & \nd & 10 pc \\
mean mass density inside $R$ & \nd & $2\times10^2$, $2\times 10^3$ & & \nd & \nd & \Msol\ \percubicpc \\
$M_{\rm dyn}(\leq R)$ & \nd & 11, 6 & & \nd & \nd & $10^8$ \Msol \\
$T_{\rm rot}$ & \nd & 3, 1 & & \nd & \nd & Myr
\enddata
\tabletypesize{\small}
\tablecomments{ The first value is for the east nucleus (or nuclear disk) and the second for the west 
when a column has two values separated with a comma.
Flux calibration uncertainties are 15\%.}
\tablenotetext{a}{Each sideband has 0.73 GHz.}
\tablenotetext{b}{Ratio of CO integrated brightness temperatures.  The CO(2--1) data are from \citest{Sakamoto99}.
The minimum baseline and resolution are matched in both lines at 41 k$\lambda$ and 0\farcs57, respectively. 
The uncertainty of the ratio is 18\%.}
\tablenotetext{c}{Integrated within 0\farcs45 from each nucleus.}
\tablenotetext{d}{The major and minor axes in FHWM and the position angle of the major axis from Gaussian fitting.
These deconvolved sizes still include the seeing disk, whose FWHM is estimated to be 0\farcs09 (\S \ref{s.red.seeing}).}
\tablenotetext{e}{Spectral index, defined with $S_\nu \propto \nu^\alpha$, between 230 GHz and 345 GHz. 
See \S \ref{s.result.continuum.Tb_spindex}.}
\end{deluxetable}

\begin{deluxetable}{rc}
\tablewidth{0pt}
\tablecaption{Continuum Positions in J2000.0 \label{t.positions} }
\tablehead{ 
	\colhead{parameter}       &
	\colhead{value}  
}
\startdata
Arp 220 E, $\alpha$  & 15$^{\rm h}$34$^{\rm m}$57\fs285           \\
                    $\delta$  & \plus23\arcdeg  30\arcmin 11\farcs27      \\
Arp 220 W, $\alpha$  & 15$^{\rm h}$34$^{\rm m}$57\fs215         \\
                    $\delta$  & \plus23\arcdeg  30\arcmin 11\farcs45      \\
   E \minus W offset & ($\Delta\alpha, \Delta\delta$) = (0\farcs96, \minus0\farcs18)  \\
    E \minus W offset & 0\farcs98, p.a.=101\arcdeg 
\enddata
\tablecomments{
See \S \ref{s.red.seeing} and \S \ref{s.result.continuum.positions} for astrometric errors.
}
\end{deluxetable}

\clearpage

\begin{deluxetable}{ccccc}
\tablecolumns{5}
\tablewidth{0pt}
\tablecaption{Opacity toward the nuclei \label{t.opacity} }
\tablehead{ 
	\colhead{$\beta$}       &
	\multicolumn{2}{c}{$\tau_{860}$}  &
	\multicolumn{2}{c}{$\tau_{1300}$} 	
\\
	\colhead{(1)}       &	
	\colhead{(2)}       &
	\colhead{(3)}       &	
	\colhead{(4)}       &
	\colhead{(5)}       			
}
\startdata
\sidehead{$R_{860/1300}$(dust, W, 0\farcs5)$=1.40 \pm 0.30$ :} 
2 & 2.3 & (1.2--5.3) & 1.0 & (0.5--2.3) \\
1 & 0.4 & ($\le 3.0$) & 0.3 & ($\le 2.0$) \\
\sidehead{$R_{860/1300}$(dust, W, 0\farcs23--0\farcs3)$=1.57 \pm 0.32$ :} 
2 & 1.7 & (0.8--3.5) & 0.7 & (0.3--1.5) \\
1 & \nd\tnm{a} & ($\le 1.4$) & \nd\tnm{a} & ($\le 0.9$) \\
\sidehead{$R_{860/1300}$(dust, E, 0\farcs5)$=1.34 \pm 0.36$ :} 
2 & 2.8 & ($\ge 1.3$) & 1.2 & ($\ge 0.6$) \\
1 & 0.8 & (\nd\tnm{b}) & 0.6 & (\nd\tnm{b}) 
\enddata
\tablecomments{
(1) Power-law index of dust emissivity ($\propto \lambda^{-\beta}$).
(2) Formal solution of 860 \micron\ optical depth from Eq. (\ref{eq.R_mm}).
(3) Range of $\tau_{860}$ corresponding to the $\pm1 \sigma$ uncertainty of $R_{860/1300}$.
(4) and (5) Same as (2) and (3), respectively, but for 1300 \micron\ optical depth. 
See \S \ref{s.result.continuum.tau_Td} for detail.
}
\tablenotetext{a}{No solution is found.}
\tablenotetext{b}{Any value is possible.}
\end{deluxetable}

\begin{deluxetable}{cccc}
\tablewidth{0pt}
\tablecaption{Upper limit to Dust temperature $T_{860}$ in the west nucleus \label{t.temperature} }
\tablehead{ 
	\colhead{$\tau_{\rm 25,fg}$}       &
	\colhead{$\theta_{25}=\theta_{860}$}       &	
	\colhead{$\theta_{25}=0\farcs25$}       &	
	\colhead{$\theta_{25}=0\farcs39$}  
\\
	\colhead{ }       &
	\colhead{$T_{25}=T_{860}$}       &	
	\colhead{$T_{25}=0.76T_{860}$}       &	
	\colhead{$T_{25}=0.61T_{860}$}  
\\
	\colhead{(1)}       &	
	\colhead{(2)}       &
	\colhead{(3)}       &	
	\colhead{(4)}       			
}
\startdata
1 & $1.3\times10^2$ & $1.4\times10^2$ & $1.5\times10^2$ \\
2 & $1.8\times10^2$ & $1.8\times10^2$ & $1.8\times10^2$
\enddata
\tablecomments{
Upper limits to the temperature of the 860-\micron\ emitting dust ($T_{860}$) in the west nucleus,
calculated in various possible cases. The unit is kelvin.
See \S \ref{s.result.continuum.tau_Td} for detail.
(1) Foreground extinction toward the nucleus in 24.5 \micron.
(2) For a less likely case in which the photosphere  is the same for 25 \micron\ and 860 \micron.
(3) and (4) For cases in which the 25 \micron\ photosphere of size $\theta_{25}$
is heated exclusively by the compact 860 \micron\ core of size $\theta_{860} \sim 0\farcs15$.
The two 25 \micron\ sizes correspond to the range set by \citest{Soifer99}. 
The temperature upper-limits will be lower if there is additional heating source 
between the 25 \micron\ photosphere and the  860 \micron\ core.
}
\end{deluxetable}

\clearpage

\begin{figure}[t]
\epsscale{0.32}
\plotone{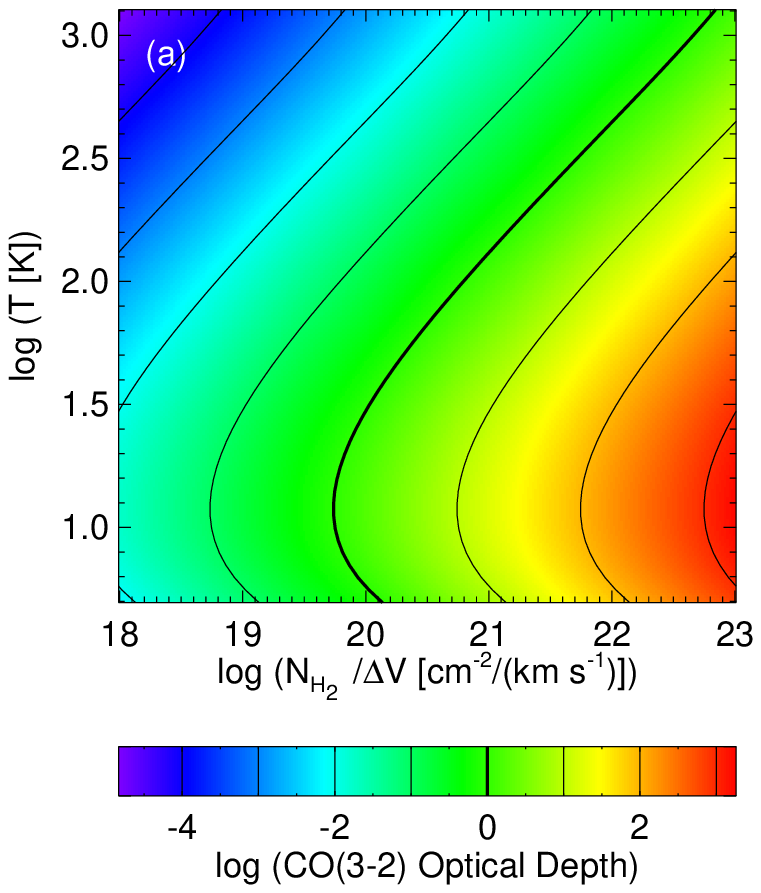} 
\plotone{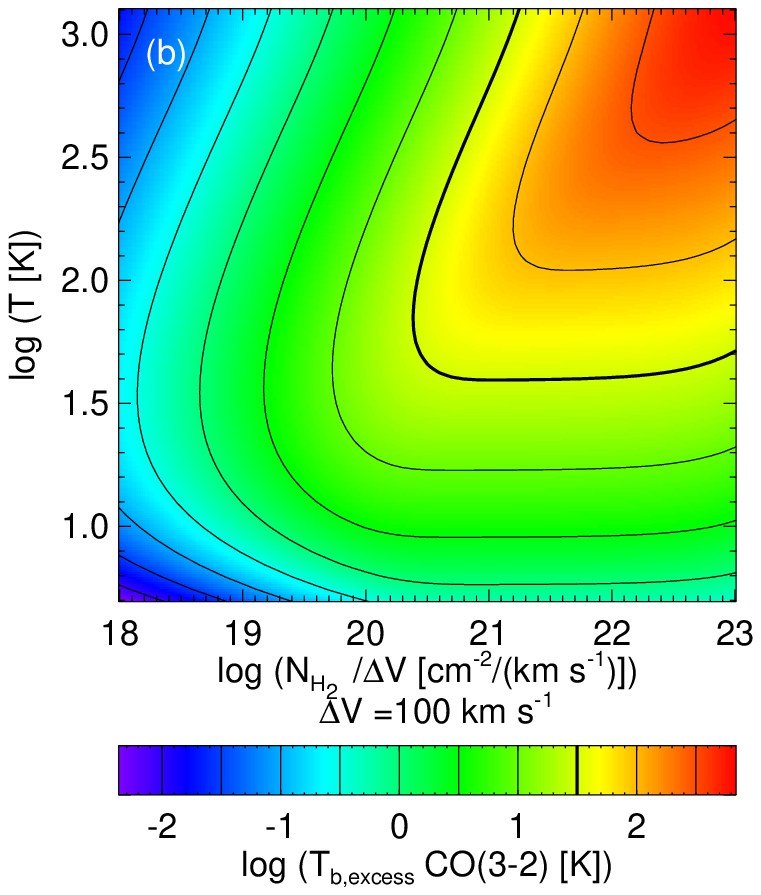} 
\plotone{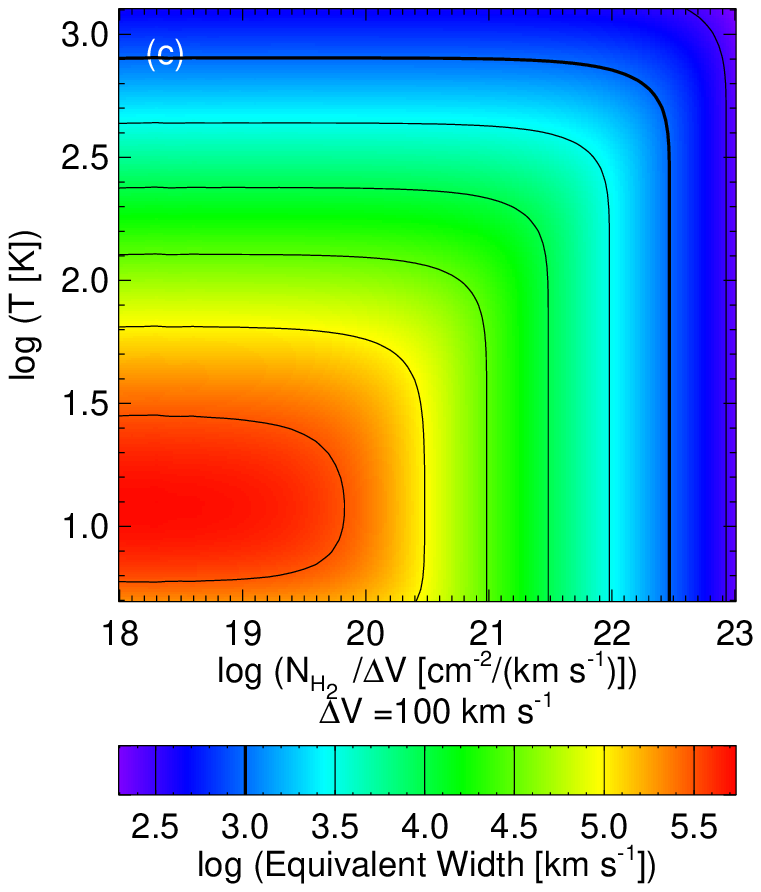} 
\epsscale{1.0}
\caption{ \label{f.lte}
CO(3--2) LTE model calculation.
(a) Optical depth,
(b) excess brightness temperature, and
(c) equivalent width with respect to thermal dust emission.
See text for the model assumptions.
The thick contours in the plot are to guide eyes and are for 
$\tau_{\rm CO(3-2)}=1$, 
$T_{\rm b,\, excess}^{\rm CO(3-2)}=30$ \KRJ, and
CO(3--2) equivalent width of $10^{3}$ \kms, 
respectively.
The dust opacity $\taud$ is 1 at $N_{\rm H_{2}}=10^{25.5}$ \persquarecm\ for the parameters
used. 
The panels (b) and (c) use an arbitrary $\Delta V$ of 100 \kms, 
but its choice has little effect on the estimate of $N_{\rm H_{2}}/\Delta V$ in Arp 220 
(see Appendix \ref{a.equiv-width}).
}
\end{figure}

\clearpage
\begin{figure}[t]
\epsscale{1.0}
\plotone{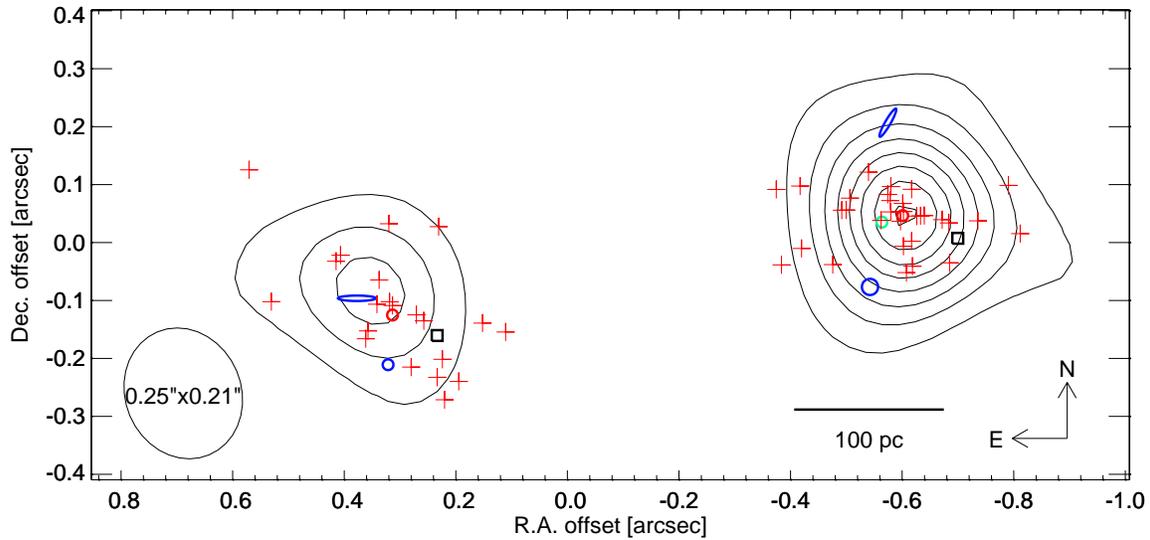}  
\caption{ \label{f.vlbisources}
Compact cm-wave sources and sub-mm continuum emission registered with it. 
Red plus signs are compact ($< 1$ pc) continuum sources thought to be  
radio supernovae or young (30--50 yr) supernova remnants \citep{Lonsdale06, Parra07}.
The red circle in each nucleus indicates the `center' (= median position) of the compact sources.
The blue circles and ellipses are OH megamasers \citep{Lonsdale98}.
The green circle is the peak of diffuse cm emission of the west nucleus, 
as measured with respect to the compact sources by \citet{Rovilos03}.
The contour map is from our 860 \micron\ continuum data, and is shifted so that the west peak
coincides with the `center' of the supernovae distribution in the nucleus.
The black squares show the positions of the sub-mm peaks before the shift.
The coordinates of the black squares and red circles are in Table \ref{t.positions} and 
\S \ref{s.result.continuum.positions}, respectively.
}
\end{figure}

\begin{figure}[t]
\epsscale{1.0}
\plottwo{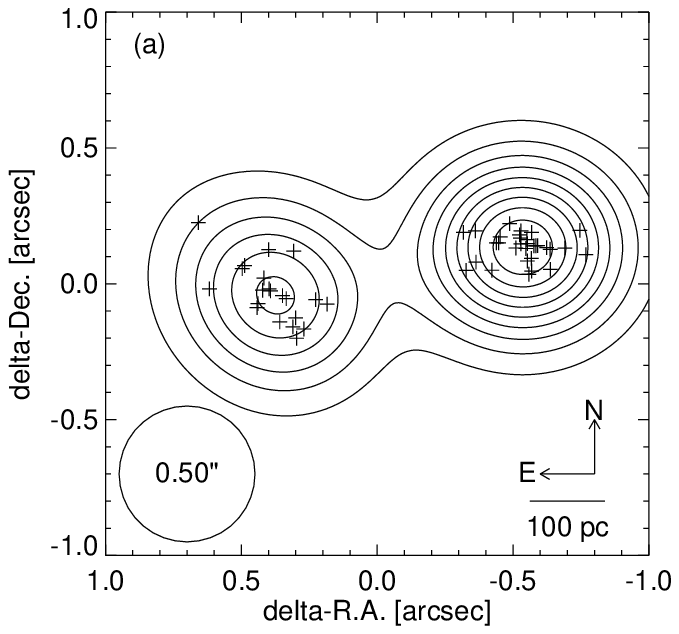}{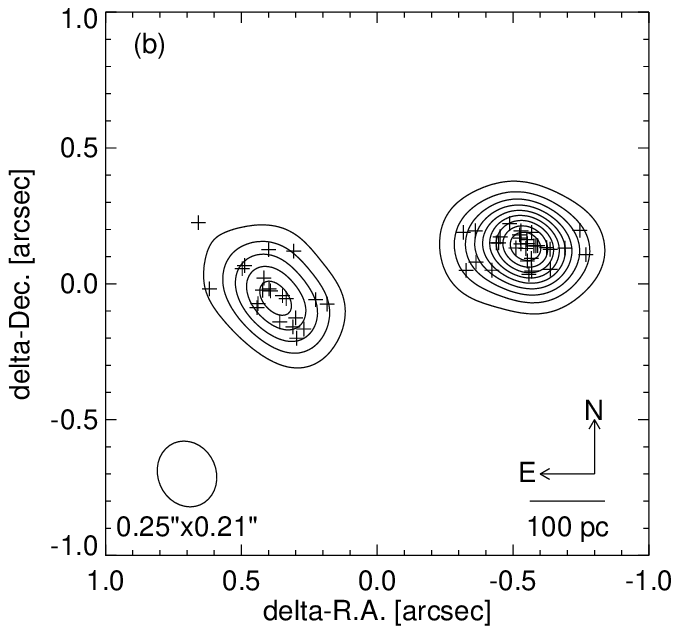}  \\ 
\caption{ \label{f.smoothedSNe}
Spatially smoothed distribution of the supernovae and supernova remnants in Arp 220.
Each plus sign is a compact ($< 1$ pc) cm continuum source
listed by \citet{Lonsdale06} or \citet{Parra07} as a radio supernova or
a young (30--50 yr) supernova remnant. 
The contour map is made by convolving the point sources with a Gaussian kernel whose
FWHM is shown at the bottom-left corner.
All sources are given the same weight to show smoothed spatial distribution of source density
and of massive star formation rate.
Contours are in 10\% steps starting from 10\% of the maximum.
}
\end{figure}

\clearpage
\appendix
\section{1. disk-gaussian deconvolution}
\label{a.disk_deconvolution}
One sometimes needs to estimate the size of a uniform-brightness source from its marginally
resolved map.
We here note that the diameter $d$ of such a disk source is about
1.6 times the FWHM one would obtain from the usual Gaussian-source deconvolution,
in which the well-known quadrature relation holds
\begin{equation}
	\label{eq.gaussian_deconvolution}
	\theta_{\rm map}^2 = \theta_{\rm int}^2+ \theta_{\rm PSF}^2,
\end{equation}
where $\theta_{\rm map}$, $\theta_{\rm int}$, and $\theta_{\rm PSF}$ are
the FWHM size of the source in the map, the intrinsic FWHM size of the Gaussian source,
and the FWHM of the Gaussian point spread function, respectively.

\begin{figure}[h]
\epsscale{0.4}
\plotone{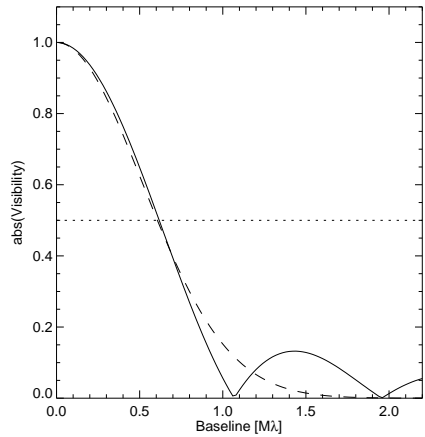}  \\
\caption{ \label{f.visamp.gauss-disk}
Visibility amplitudes of a Gaussian of $\theta_{\rm FWHM}=0\farcs15$ (dashed line) 
and a disk with a diameter $d=1.57\theta_{\rm FWHM}$  (solid line).
}
\end{figure}

A uniform-brightness circular source of the diameter $d$ and total flux 1 has the visibility 
\begin{equation}
	D(\rho)
	=
	\frac{2J_1(\pi d \rho)}{\pi d \rho},
\end{equation}
where $\rho$ is the baseline length and $J$ is the Bessel function of the first kind.
A Gaussian source of FWHM $\theta_{\rm int}$ and the same total flux has the visibility
\begin{equation}
	G(\rho)
	=
	\exp
	\left(
		- \frac{\pi^2 \theta_{\rm int}^2 \rho^2}{4\ln 2}
	\right).
\end{equation}
The half-power $uv$ radius of the visibility functions is at $\pi d \rho \approx 2.2151$ for the disk
and $\pi \theta_{\rm int} \rho = 2 \ln 2$ for the Gaussian. 
The two functions agree at their half power points if
\begin{equation}
	\label{eq.d-theta.vis}
	d \approx \frac{2.2151}{2\ln 2} \, \theta_{\rm int} \approx 1.60\, \theta_{\rm int}.
\end{equation}
Fig. \ref{f.visamp.gauss-disk} shows that the two visibility functions resemble each other
out to longer baselines when the diameter is slightly smaller than this. 
(The figure is for $d=1.57 \, \theta_{\rm int}$.)

Taking the convolution of the disk with the PSF more directly, the convolution relation
in the $uv$ domain is
\begin{equation}
	\exp
	\left(
		- \frac{\pi^2 \theta_{\rm map}^2 \rho^2}{4\ln 2}
	\right)
	\approx
	\frac{2J_1(\pi d \rho)}{\pi d \rho}
	\exp
	\left(
		- \frac{\pi^2 \theta_{\rm PSF}^2 \rho^2}{4\ln 2}
	\right).		
\end{equation}
Requiring this to be exact at the half-power point 
(i.e., at $\rho$ that satisfies $\pi \theta_{\rm map} \rho = 2 \ln 2$) one obtains
\begin{equation}
	\theta_{\rm map}^2 = \frac{\ln 2}{2}d^2+ \theta_{\rm PSF}^2
\end{equation}	
for $d/\theta_{\rm map} \ll 1$. 
Comparison of this with \ref{eq.gaussian_deconvolution} gives 
a similar coefficient as in \ref{eq.d-theta.vis},
\begin{equation}
	\label{eq.d-theta.im}
	d
	\approx
	1.70 \, \theta_{\rm int} .
\end{equation}

\section{2. equivalent width of molecular line with respect to dust continuum}
\label{a.equiv-width}
The equivalent width of a molecular line to the dust continuum emission
for an isothermal mixture of molecular gas and dust has the following
relation to the ISM properties.
Let  \taum\ and \taud\ be the optical depths of the molecular line 
and the dust continuum adjacent to the line's frequency $\nu$, respectively.
The total optical depth at the line frequency is $\taum + \taud$.
The emission intensities on and off the line frequency are, for LTE excitation at the gas (and dust)
temperature $T$, 
\begin{eqnarray}
	I_{\nu}^{\rm line+cont.}
	&=&
	B_{\nu}(T)
	\left[ 1 - e^{-(\taum+\taud)} \right] 
	+
	B_{\nu}(\Tcmb)
	e^{-(\taum+\taud)}
	\\
	I_{\nu}^{\rm cont.}
	&=&
	B_{\nu}(T)
	\left[ 1 - e^{-\taud} \right] 
	+
	B_{\nu}(\Tcmb)
	e^{-\taud},		
\end{eqnarray}
where $B_{\nu}$ is the Planck function
and \Tcmb\ is the cosmic background temperature.
The excess brightness temperature of the dust continuum is
the excess of intensity, expressed in the Rayleigh-Jeans temperature, over an off-source position.
\begin{eqnarray}
	\Tcont
	&=&
	\frac{c^{2}}{2k\nu^{2}}
	\left[  
		I_{\nu}^{\rm cont.}(\taud) - 	I_{\nu}^{\rm cont.}(0)
	\right]
	=
	\left( 1 - e^{-\taud} \right)
	\left[ J(\nu, T) - J(\nu, \Tcmb) \right],
\end{eqnarray}
where $J(\nu, T) \equiv \frac{h\nu}{k} (e^{h\nu/kT}-1)^{-1}$ is the effective radiation temperature
and $c$,$k$, and $h$ are the speed of light, Boltzmann, and Planck constants, respectively. 
The excess brightness temperature of the line after continuum subtraction is,
\begin{eqnarray}
	\Tline
	&=&
	\frac{c^{2}}{2k\nu^{2}}
	\left\{
	\left[  I_{\nu}^{\rm line+cont.}(\taum+\taud) - I_{\nu}^{\rm line+cont.}(0) \right]	
	-
	\left[  I_{\nu}^{\rm cont.}(\taud) - I_{\nu}^{\rm cont.}(0) \right]
	\right\}
	\nonumber \\
	& = &
	e^{-\taud} \left( 1 - e^{-\taum} \right)
	\left[ J(\nu, T) - J(\nu, \Tcmb) \right].
\end{eqnarray}
Hence,  the equivalent width for a line with a velocity width of \DeltaV, is
\begin{equation}	\label{eq.weq-formal}
	\Weq 
	\equiv
	\frac{\Tline \DeltaV}{\Tcont} 
	= 
	\frac{1- e^{-\taum}}{e^{\taud}-1} \DeltaV.
\end{equation}
The beam filling factor of the ISM, which should be multiplied to
\Tline\ and \Tcont\ to compare them with observations, cancels out in \Weq.

The line optical depth of a CO-like molecule in LTE is written in the following form
\begin{equation}
	\taum \simeq a \frac{\Nhh}{\DeltaV} T^{-2}
\end{equation}
at high temperatures ($T \gg h\nu/k$),
and the dust optical depth can be written as
\begin{equation}
	\taud = b \Nhh,
\end{equation}
where \Nhh\ is the \HH\ column density, 
$a$ is a constant that reflects the properties of the molecule in question
and its abundance with respect to \HH, 
and $b$ is another constant that contains gas-to-dust mass ratio and the dust mass opacity coefficient.
The equivalent width can be uniquely calculated on the $\Nhh/\DeltaV$ -- $T$ plane
when the dust emission is optically thin (i.e. $e^{\taud}-1 \approx \taud$).
Moreover, if $\taud \ll 1$ then 
\Weq\ depends on only one of the two parameters except when the molecular line 
has a moderate opacity.
\begin{equation}	\label{eq.weq-approx}
	\Weq 
	\approx
	\left\{
	\begin{array}{ll}
	(b \Nhh/\DeltaV)^{-1}  & \quad \mbox{(for $\taud \ll 1$ and $\taum \gg 1$)} \\
	a b^{-1} T^{-2}   & \quad \mbox{(for $\taud \ll 1$ and $\taum \ll 1$)}	
	\end{array}
	\right.
\end{equation}
This is why the iso-\Weq\ contours are mostly parallel to either the
$\Nhh/\DeltaV$ or $T$ axis in Fig. \ref{f.lte} (c) and why \Weq\ can be used to 
estimate one of the two parameters.
The $\Nhh/\DeltaV$ estimated using the approximate formula (\ref{eq.weq-approx})  overestimates 
the correct one from (\ref{eq.weq-formal}) 
by a factor of $\frac{\Delta V}{W_{\rm eq}}/\ln (1+ \frac{\Delta V}{W_{\rm eq}})$, 
or by less than 14\% for $\DeltaV \leq 300$ \kms\ and $\Weq = 10^{3}$ \kms.
Finally, the factors that can bias the above evaluation include
non-LTE excitation, disagreement of gas and dust temperatures, 
non-uniform ISM density, column density, and temperature, 
and
variations of molecular abundance and gas-to-dust ratio.

\section{3. dust continuum intensity and SN surface density }
\label{a.Id-Sigma_sn}
Surface density of supernovae and supernova remnants (called SN in short), 
$\Sigma_{\rm SN}$, must be approximately proportional
to the intensity of dust continuum, $I_{\rm d}$, for a dusty starburst region 
if some plausible assumptions hold.
First, if the number of SN ($N_{\rm SN}$) in a region  is proportional to the star formation rate
$R_{\rm SF}$ in the region then we have
\begin{eqnarray}
	N_{\rm SN} & \propto & R_{\rm SF}	\\
	\therefore \quad
	\Sigma_{\rm SN} & \propto & \Sigma_{\rm SF}.
\end{eqnarray}
The second formula expresses the first one in terms of surface density.
Secondly, assuming that the region is dominantly heated by star formation (of luminosity $L_{\rm SF}$) 
and cooled by dust emission, we have
\begin{eqnarray}
	L_{\rm SF} &  \propto & T^{4}l^{2}	\\
	\therefore \quad
	\Sigma_{\rm SF} & \propto & T^{4},
\end{eqnarray} 
where $l$ and $T$ are the size and temperature of  the region, respectively.
Thirdly, assuming that star formation surface density is related to the
gas surface density $\Sigma_{\rm g}$, we employ the Schmidt-Kennicutt law
\begin{eqnarray}
	\Sigma_{\rm SF} & \propto & \Sigma_{\rm g}^{\;\; n},
\end{eqnarray}
where $n$ is widely thought to be in the range of 1--2 \citep{Kennicutt98}.
Finally, if the dust emission that we observe in (sub)millimeter
is of moderate optical depth or optically thin
(i.e., $\tau_{\rm d} \lesssim 1$ for dust opacity) then we have
\begin{equation}
	I_{\rm d} \propto \tau_{\rm d}T \propto \Sigma_{\rm g} T.
\end{equation}
It follows from these relations that
\begin{equation}
	T \propto \Sigma_{\rm SN}^{\;\;\;\;\; 1/4}
\end{equation}
and
\begin{equation}
	I_{\rm d} \propto \Sigma_{\rm SN}^{\;\;\;\;\; \frac{1}{n}+\frac{1}{4}}.
\end{equation}
The exponent $\frac{1}{n}+\frac{1}{4}$ is in the range of 0.75--1.25 for $n = $1--2 
and is 0.96 for the canonical $n= 1.4$.

\end{document}